\documentclass[11pt]{article}
\usepackage{graphicx}
\usepackage{amsmath}
\usepackage{amssymb}
\usepackage{amsxtra}

\usepackage{CJK}

\usepackage{graphicx,amsmath,amssymb}

\usepackage{graphicx}
 \usepackage{epstopdf}
\usepackage{amsmath}
\usepackage{dcolumn}% Align table columns on decimal point
\usepackage{bm}% bold math
\usepackage{slashed}
\usepackage{siunitx}
\usepackage{ulem,xpatch}
\usepackage{hyperref}% add hypertext capabilities
\usepackage[mathlines]{lineno}
\usepackage{comment}
\usepackage{soul}
\usepackage{xcolor}
\usepackage{xfrac}
\hypersetup{colorlinks, linkcolor = [rgb]{0,0.0,0.5}, citecolor = [rgb]{0,0.0,0.5}, urlcolor = [rgb]{0,0.0,0.5}}

\title{Supersymmetric and Other Novel Features 
of Hadron Physics from Light-Front Holography
}

\author{Stanley~J.~Brodsky\\SLAC National Accelerator Laboratory, Stanford University \\e-mail: sjbth@slac.stanford.edu}
\begin{document}
\maketitle

\begin{abstract}
I survey recent developments in hadron physics which follow from the application of superconformal quantum
mechanics and light-front holography. This includes new insights into the
physics of color confinement, chiral symmetry, the spectroscopy and
dynamics of hadrons, as well as surprising supersymmetric relations
between the masses of mesons, baryons, and tetraquarks.  I also will
discuss some novel features of QCD -- such as color transparency,
hidden color, and asymmetric intrinsic heavy-quark phenomena. The elimination of renormalization scale ambiguities and the modification of QCD sum rules due to diffractive phenomena are also briefly reviewed.
\end{abstract}

\noindent QCD, Light-Front, Holography, Intrinsic Charm, Color Transparency, Supersymmetry, Principle of Maximum Conformality

% optionally
%\noindent PACS: ... list of PACS codes

\section{Color Confinement and Light-Front Holography}

A key problem in hadron physics is to obtain a first approximation to QCD which can accurately predict not only the spectroscopy of hadrons, but also the light-front wave functions which underly their properties and dynamics.
Guy de T\'eramond, Guenter Dosch, and I~\cite{Brodsky:2013ar}
have shown that a mass gap and a fundamental color confinement scale can be derived from light-front holography -- the duality between five-dimensional anti-de Sitter (AdS) space physical 3+1 spacetime using light-front time.  The combination of superconformal quantum mechanics~\cite{deAlfaro:1976je, Fubini:1984hf}, light-front quantization~\cite{Dirac:1949cp} and the holographic embedding on a higher dimensional gravity theory~\cite{Maldacena:1997re} (gauge/gravity correspondence) has led to new analytic insights into the structure of hadrons and their dynamics~\cite{deTeramond:2008ht, Brodsky:2013ar, deTeramond:2014asa, Dosch:2015nwa, Brodsky:2014yha, Brodsky:2020ajy}. This new approach to nonperturbative QCD dynamics, {\it holographic light-front QCD}, has led to effective semi-classical relativistic bound-state equations for arbitrary spin~\cite{deTeramond:2013it}, and it incorporates fundamental properties which are not apparent from the QCD Lagrangian, such as the emergence of a universal hadron mass scale, the prediction of a massless pion in the chiral limit, and remarkable connections between the spectroscopy of mesons, baryons and tetraquarks across the full hadron spectrum~\cite{Dosch:2015bca, Dosch:2016zdv, Nielsen:2018uyn, Nielsen:2018ytt}.  See  Fig. \ref{Bledslides7.pdf}.

\begin{figure}
 \begin{center}
\includegraphics[height= 12cm,width=15cm]{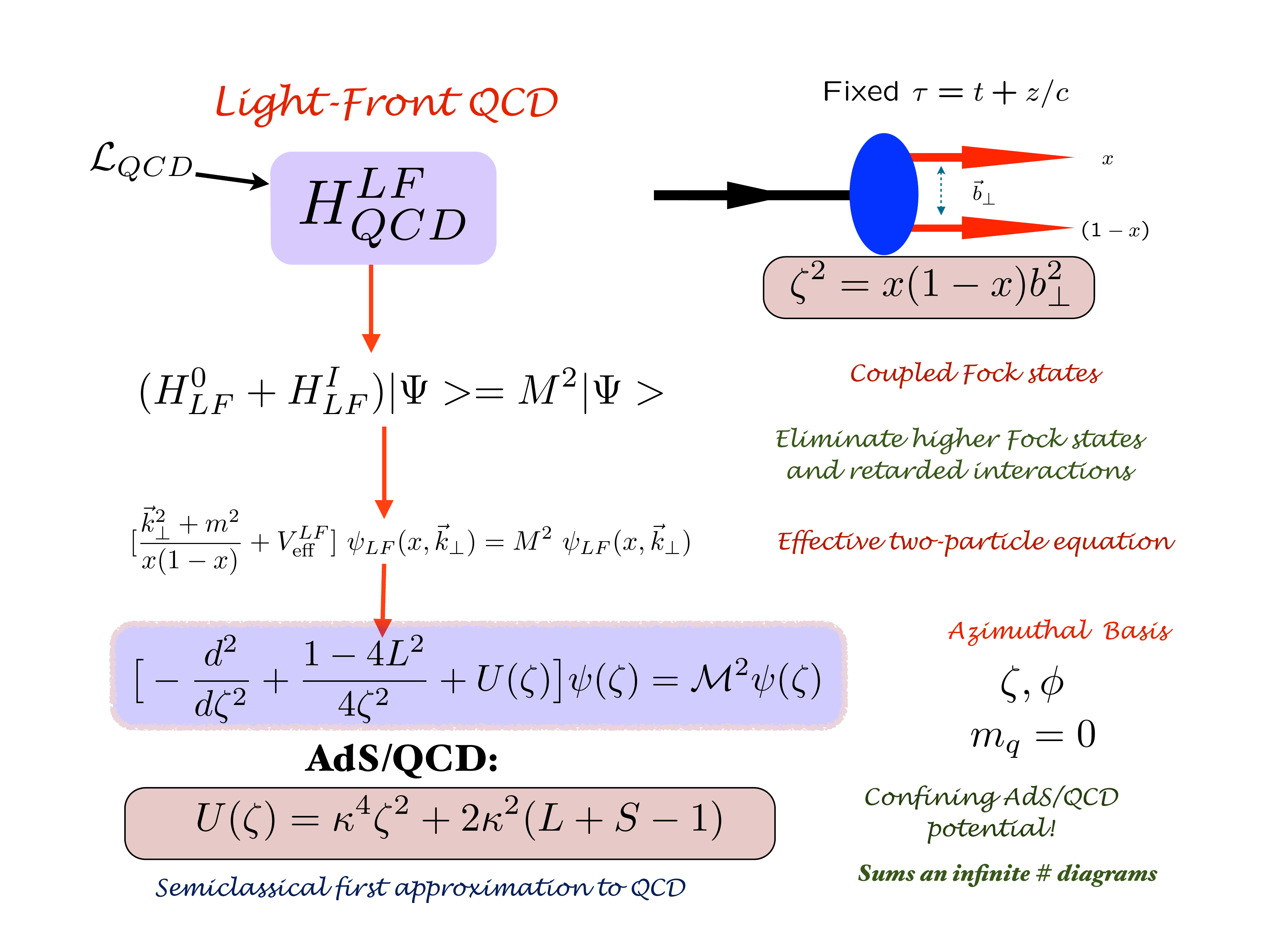}
\end{center}
\caption{Derivation of the Effective Light-Front Schr\"odinger Equation from QCD.  As in QED, one reduces the LF Heisenberg equation 
$H_{LF} |\Psi \rangle    = M^2 |\Psi \rangle $ 
to an effective two-body eigenvalue equation for $q \bar q$ mesons by systematically eliminating higher Fock states. 
One utilizes the LF radial variable $\zeta$, where $\zeta^2 = x(1-x)b^2_\perp$ is conjugate to the $q \bar q$ LF kinetic energy $k^2_\perp\over x(1-x)$ for $m_q=0$. This allows the reduction of the dynamics to a single-variable bound-state equation acting on the valence $q \bar q$ Fock state.  The confining potential $U(\zeta)$, including its spin-$J$ dependence, is derived from the soft-wall AdS/QCD model with the dilaton  $e^{+\kappa^2 z^2 },$ where $z$ is the fifth coordinate of AdS$_5$ holographically dual  to $\zeta$. See Ref.~\cite{Brodsky:2013ar}.   The resulting light-front harmonic oscillator confinement potential $\kappa^4 \zeta^2 $ for light quarks is equivalent to a linear confining potential for heavy quarks in the instant form~\cite{Trawinski:2014msa}. }
\label{FigsJlabProcFig2.pdf}
\end{figure}

The light-front  equation for mesons of arbitrary spin $J$ can be derived~\cite{deTeramond:2013it}
from the holographic mapping of  the ``soft-wall'  modification~\cite{Karch:2006pv} of AdS$_5$ space with the specific dilaton profile $e^{+\kappa^2 z^2}$,  where one identifies the fifth dimension coordinate $z$ with the light-front coordinate $\zeta$, where $ \zeta^2 = b^2_\perp x(1-x)$.  As emphasized by Maldacena~\cite{Maldacena:1997re}, a key feature of five-dimensional AdS$_5$ space is that it provides a geometrical representation of the conformal group.
Moreover AdS$_5$  is holographically dual to 3+1  spacetime where the time coordinate is light-front time $\tau = t+ z/c$.  The resulting light-front potential has the unique form of a harmonic oscillator $\kappa^4 \zeta^2$ in the 
light-front invariant variable $\zeta$. The result is  a frame-independent relativistic equation of motion for  $q \bar q$ bound states --  a ``Light-Front Schr\"odinger Equation"~\cite{deTeramond:2008ht}, analogous to the nonrelativistic radial Schr\"odinger equation in quantum mechanics.  This bound state equation  incorporates color confinement and other essential spectroscopic and dynamical features of hadron physics, including a massless pion for zero quark mass and linear Regge trajectories with the same slope in both the radial quantum number $n$   and the internal  orbital angular momentum $L$.   
The derivation of the confining Light-Front Schr\"odinger Equation is outlined in Fig. \ref{FigsJlabProcFig2.pdf}.

The predictions for hadron spectroscopy and dynamics ~\cite{deTeramond:2014asa,Dosch:2015nwa,Dosch:2015bca} include effective QCD light-front equations for both mesons and baryons based on the generalized supercharges of superconformal algebra~\cite{Fubini:1984hf}. 
The supercharges connect the baryon and meson spectra  and their Regge trajectories to each other in a remarkable manner: each meson has internal  angular momentum one unit higher than its superpartner baryon:  $L_M = L_B+1.$  See  Fig. \ref {FigsJlabProcFig3.pdf}.   Only one mass parameter $\kappa$ appears; it sets the confinement and the hadron mass scale in the  chiral limit, as well as  the length scale which underlies hadron structure.  Light-Front Holography  in fact not only predicts meson and baryon  spectroscopy  successfully, but also hadron dynamics:  light-front wave functions, vector meson electroproduction, distribution amplitudes, form factors, and valence structure functions.  The holographic duality connecting LF physics in 3+1 physical space-time with AdS space in 5 dimensions is illustrated in Fig. \ref{Bledslides4.pdf}. The dilaton $e^{\kappa z^2}$ modification of the metric of AdS space leads to a color-confining potential in the LF Schr\"odinger equation.

The  combination of light-front dynamics, its holographic mapping to AdS$_5$ space, and the de Alfaro-Fubini-Furlan (dAFF) procedure~\cite{deAlfaro:1976je} provides new  insight into the physics underlying color confinement, the nonperturbative QCD coupling, and the QCD mass scale.   A comprehensive review is given in Ref.~\cite{Brodsky:2014yha}.  The $q \bar q$ mesons and their valence LF wave functions are the eigensolutions of a frame-independent bound state equation, the  Light-Front Schr\"odinger Equation.  The mesonic $q\bar q$ bound-state eigenvalues for massless quarks have the simple Regge form $M^2(n, L, S) = 4\kappa^2(n+L +S/2)$.
The equation predicts that the pion eigenstate  $n=L=S=0$ is massless at zero quark mass.

\begin{figure}
 \begin{center}
\includegraphics[height= 8cm,width=15cm]{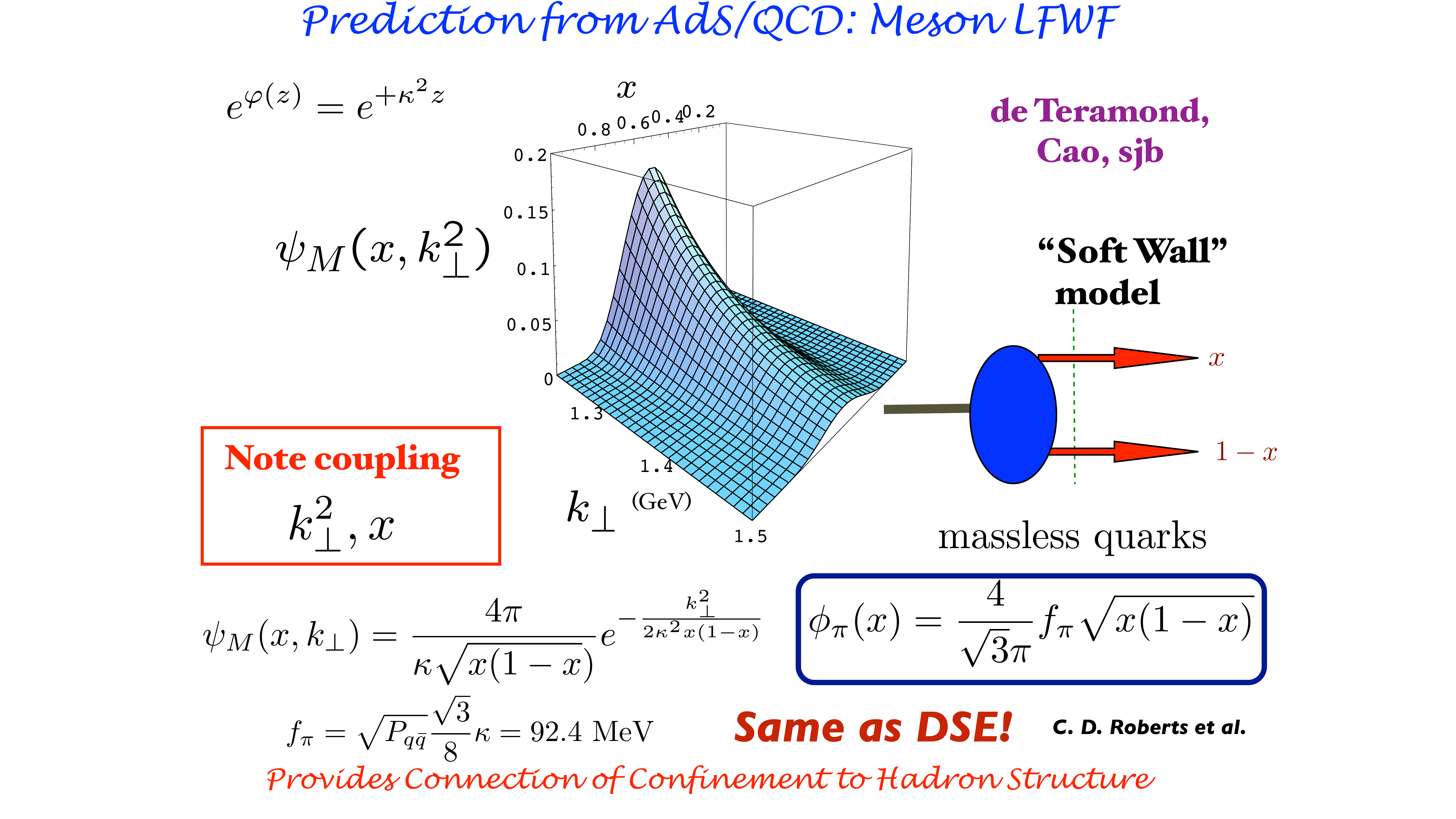}
\end{center}
\caption{The LF wave function of the pion predicted by LF holography.  The results are consistent with analyses based on the Dyson-Schwinger equation.}
\label{Bledslides6.pdf}
\end{figure}

\section{Light-Front Holography QCD and Supersymmetric Features 
of Hadron Physics
}  

\begin{figure}
 \begin{center}
\includegraphics[height= 10cm,width=15cm]{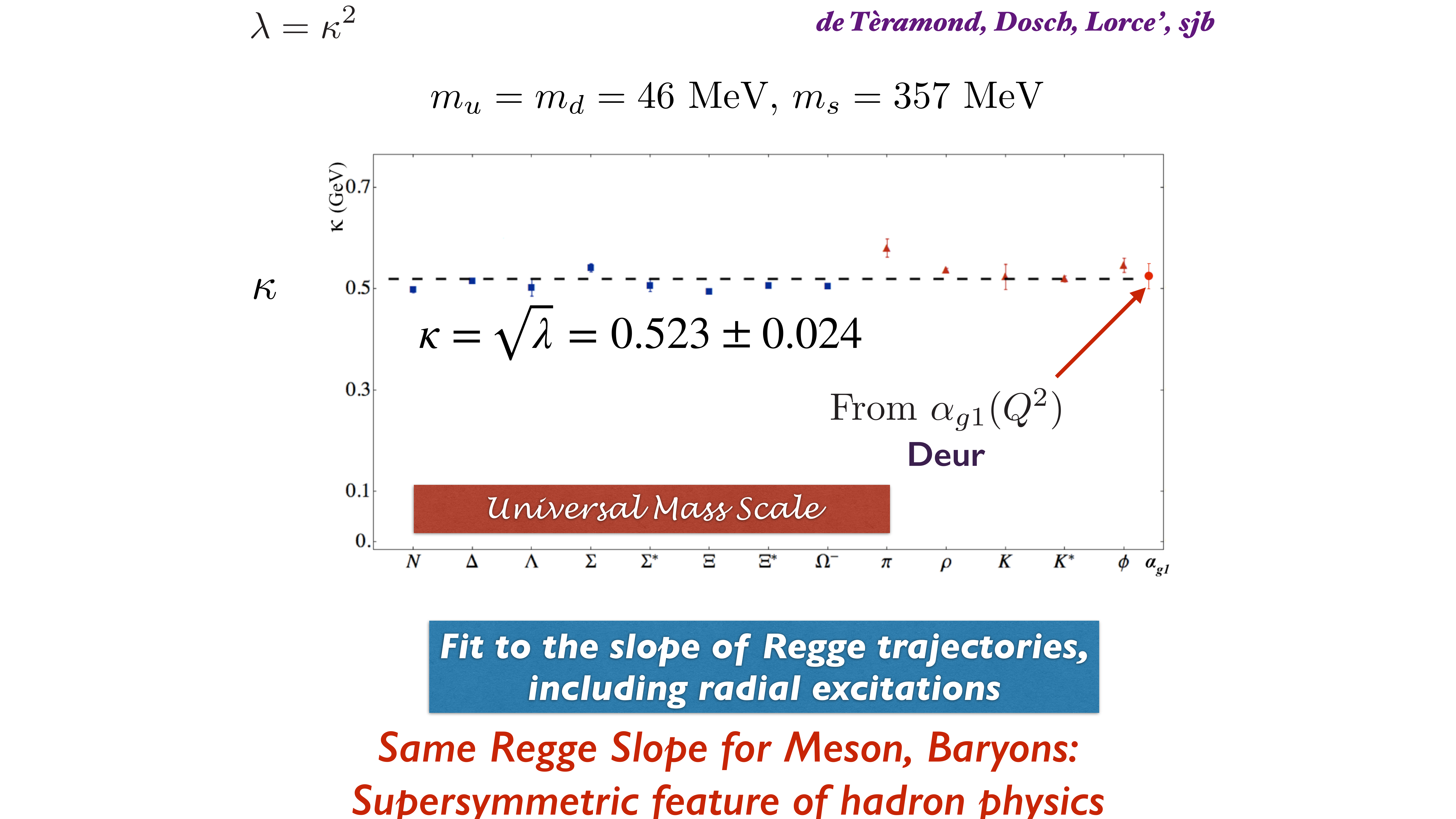}
\end{center}
\caption{The slopes of the measured meson and baryon Regge trajectories.}
\label{Bledslides2.pdf}
\end{figure}

\begin{figure}
 \begin{center}
\includegraphics[height= 9cm,width=18cm]{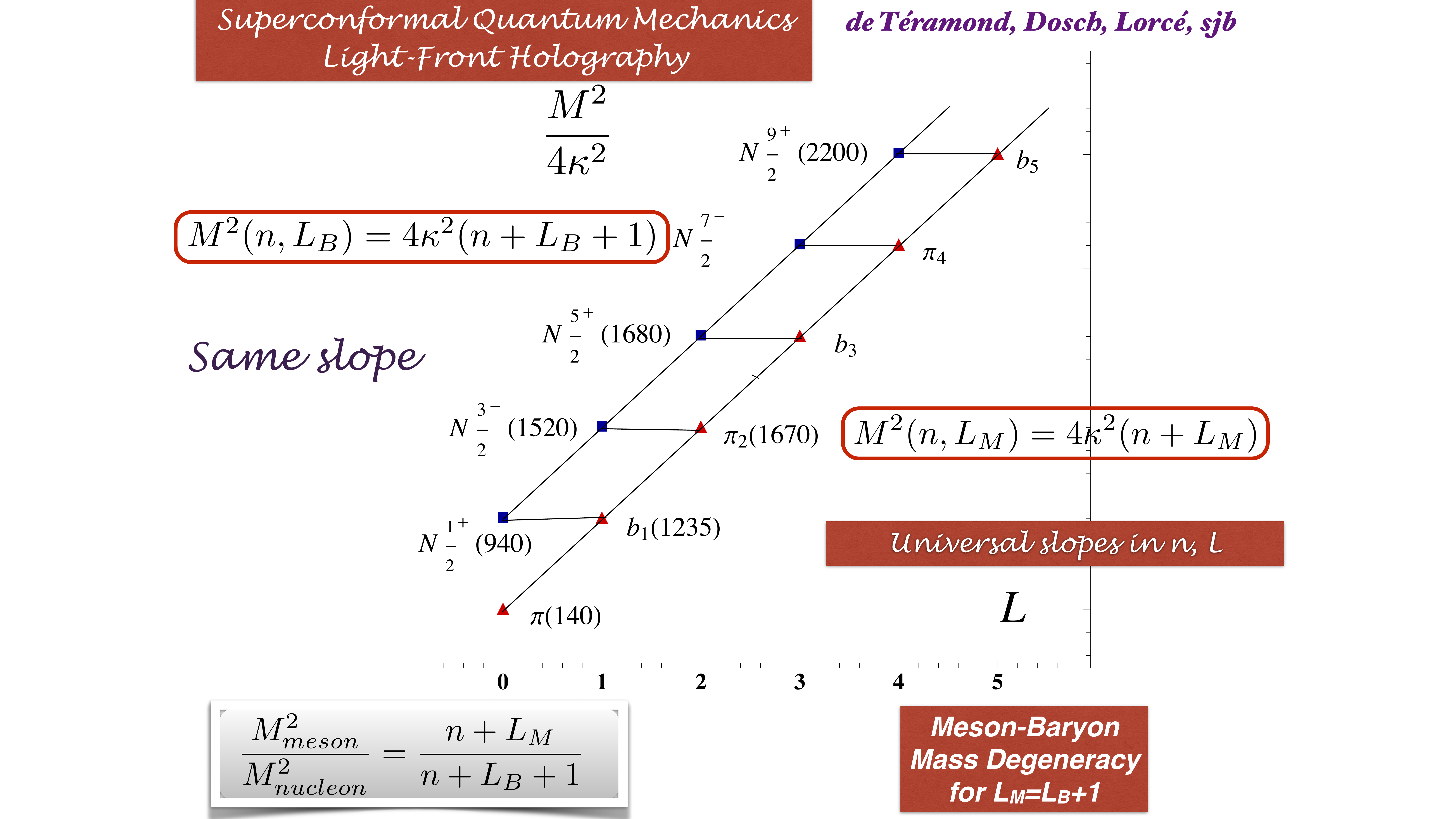}
\includegraphics[height=9 cm,width=15cm]{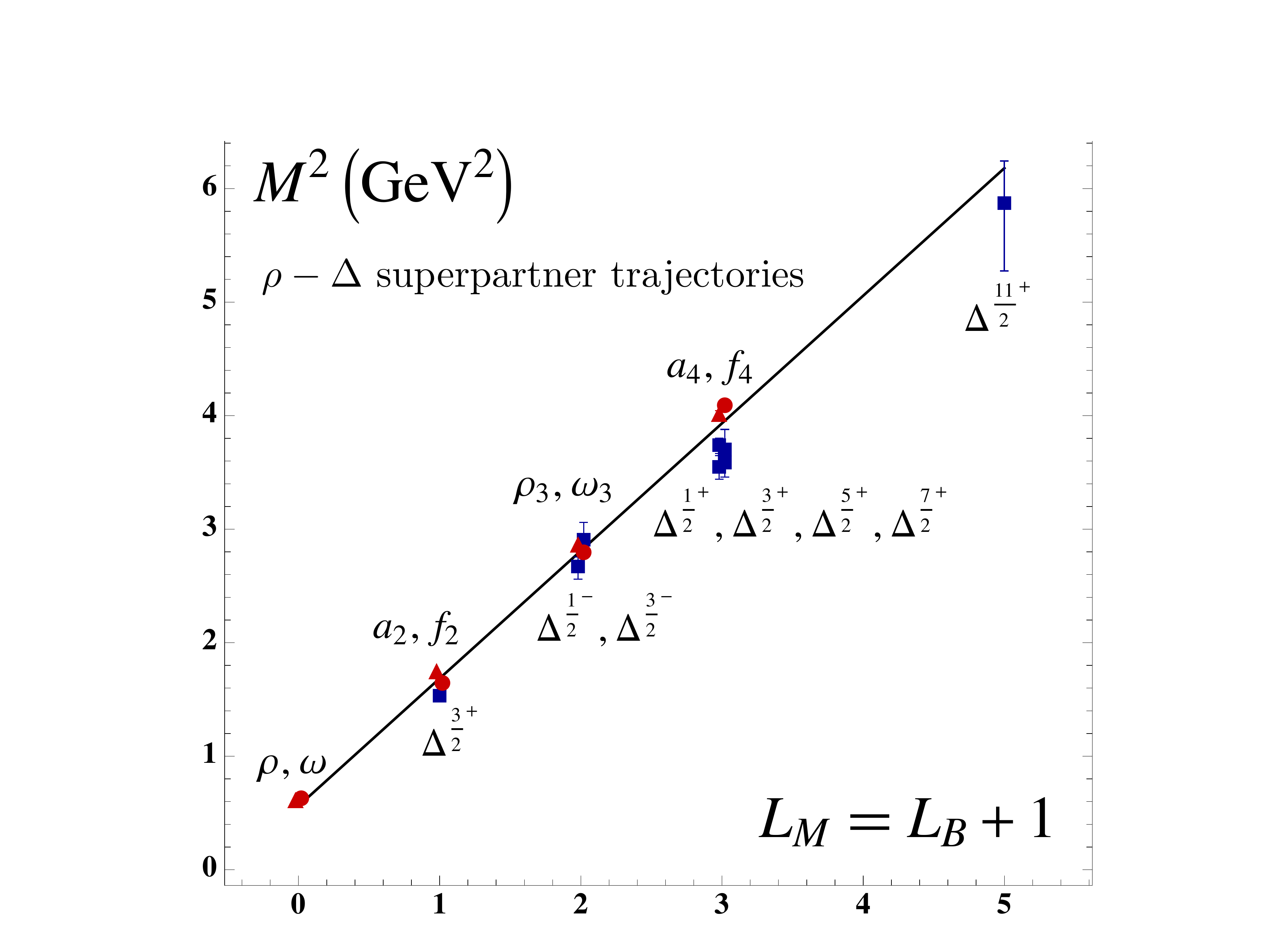}
\end{center}
\caption{Examples of supersymmetric meson and baryon Regge trajectories. Comparison of the pion  and proton trajectories and the comparison of the $\rho/\omega$ meson Regge trajectory with the $J=3/2$ $\Delta$  baryon trajectory.   The degeneracy of the  meson and baryon trajectories if one identifies a meson with internal orbital angular momentum $L_M$ with its superpartner baryon with $L_M = L_B+1$ using superconformal algebra.
See Refs.~\cite{deTeramond:2014asa,Dosch:2015nwa}.} 
\label{Bledslides8.pdf}
\end{figure}

\begin{figure}
 \begin{center}
\includegraphics[height= 8cm,width=15cm]{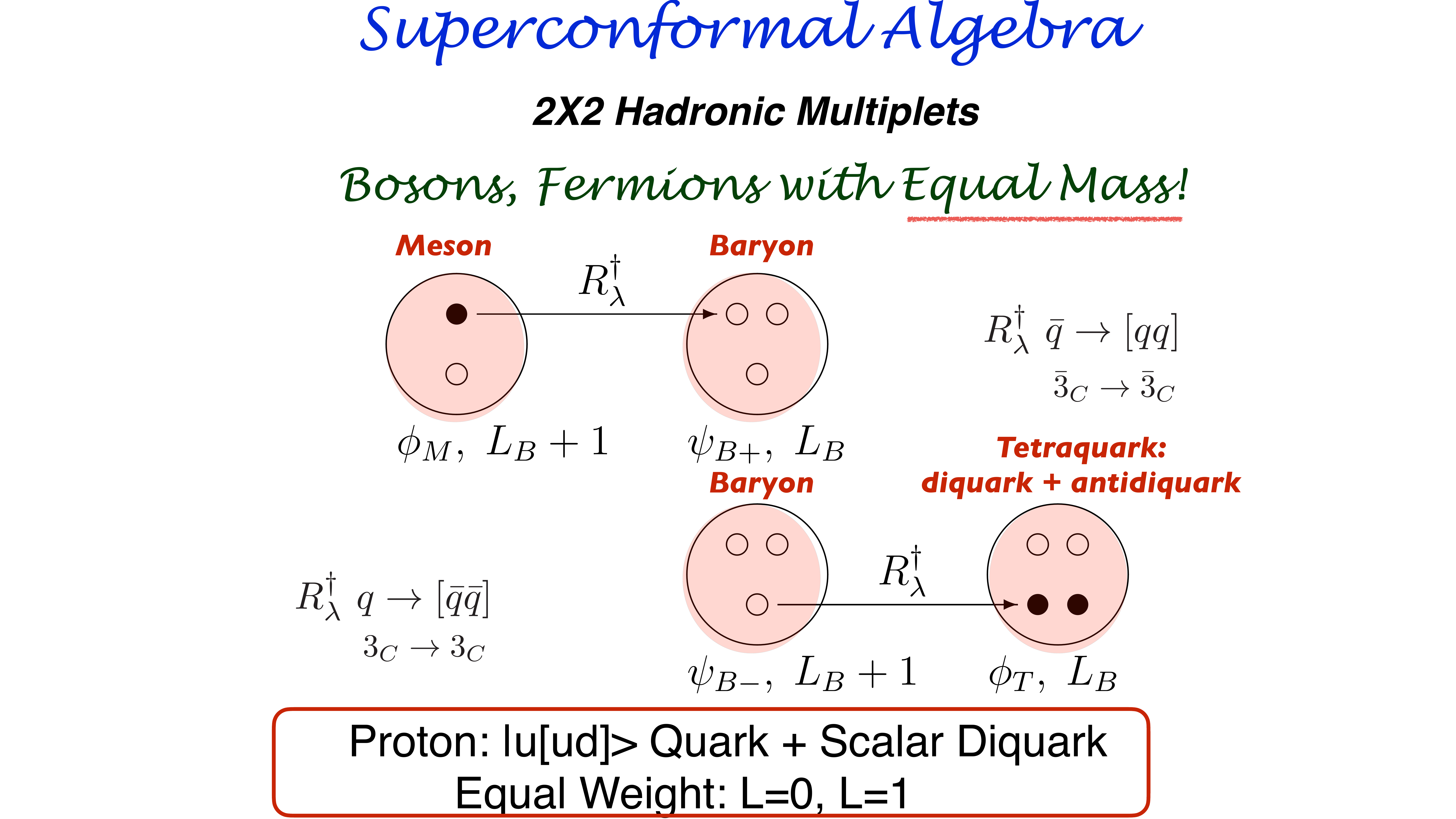}
\end{center}
\caption{The supersymmetric meson-baryon-tetraquark 4-plet.  The operator $R^\dagger_\lambda$  transforms an antiquark $\bar 3_C$ into a diquark $\bar 3_C$.}
\label{Bledslides7.pdf}
\end{figure}

\begin{figure}
 \begin{center}
\includegraphics[height= 10cm,width=15cm]{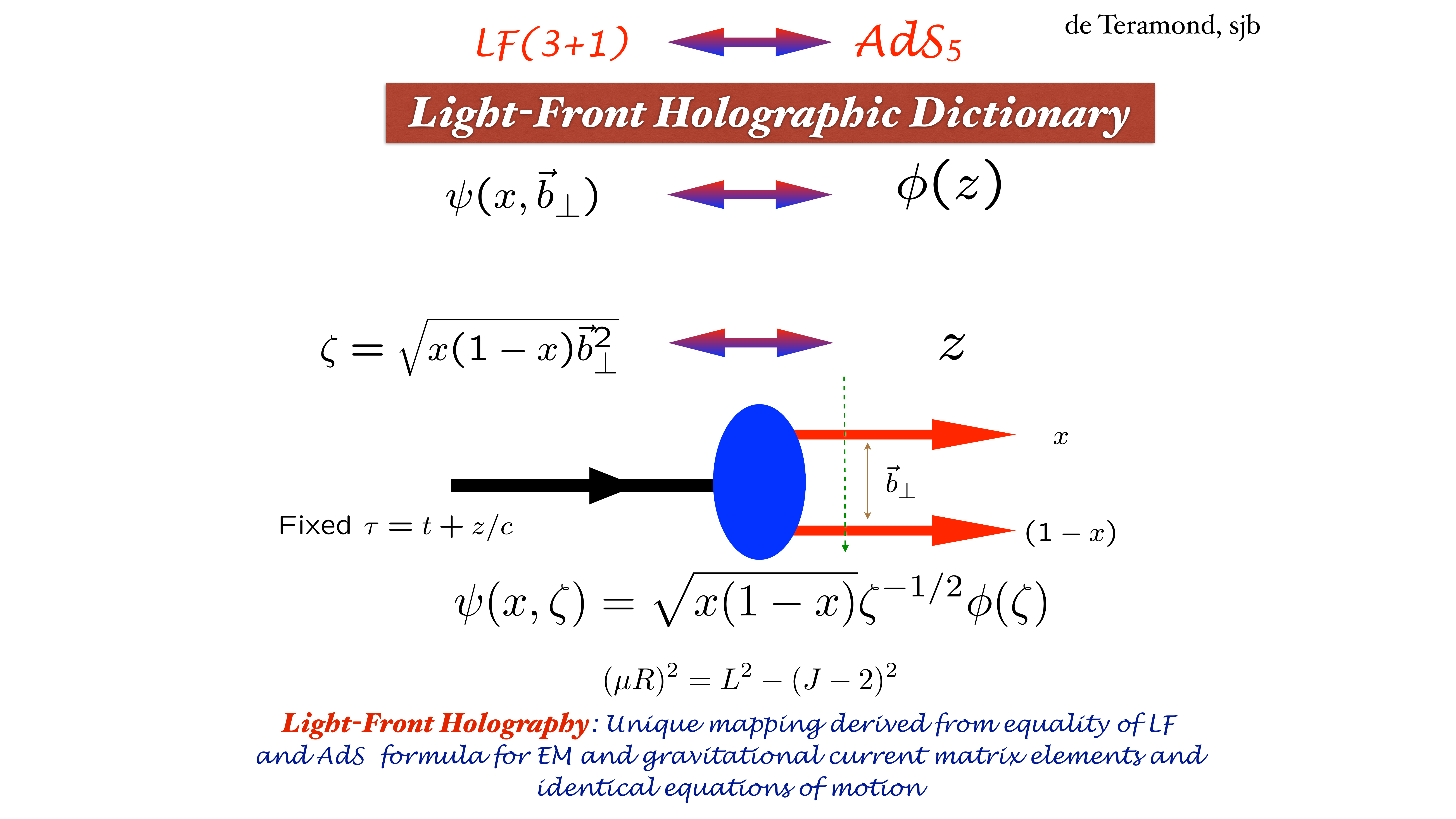}
\end{center}
\caption{The holographic duality connecting LF physics in 3+1 physical space-time with AdS space in 5 dimensions. 
The coordinate $z$ in  the fifth dimension of AdS space  is holographically dual to the LF radial variable $\zeta$ where $\zeta^2 = b^2_\perp x(1-x)$. }
\label{Bledslides4.pdf}
\end{figure}

One of the most remarkable feature of hadron spectroscopy is that, to a very good approximation, mesons and baryons are observed to lie on almost identical Regge trajectories:
$M^2_M = 2 \kappa^2 (n + L_M) $ 
for mesons with light quarks and  
$M^2_B = 2 \kappa^2 (n + L_B +1) $ 
for baryons with light quarks. 
The slopes $\lambda =\kappa^2$  in $M^2_H(n,L)$ are identical for both mesons and baryons in both the principal number $n$ and orbital angular momentum  $L$. 
(The index $n$ can be interpreted as the number of nodes in the resulting two-body wave function. )
The universality of the slopes of Regge trajectories across the hadronic spectrum is shown in Fig. \ref{Bledslides2.pdf}.  
An example comparing the pion and proton trajectories is shown in Fig. \ref{Bledslides8.pdf}.
This degeneracy between the Regge slopes of the two-body mesons and three-body baryons provides compelling evidence that two of the three quarks in the baryon valence Fock state pair up as 
diquark clusters. Then $L_M$ represents the orbital and angular momentum between the $3_C$ quark and  $\bar 3_C$ antiquark for mesons, and 
$L_B$ represents the orbital angular momentum between the $3_C$ quark and a $\bar 3_C$ spin-0 $[qq]$  or spin-1 $(qq)$ diquark in baryons.  
The identical $3_C- \bar 3_C$ color-confining interaction appears for mesons and baryon.  The index $n$ can be interpreted as the number of nodes in the resulting two-body wave function.

The unified spectroscopy of hadronic bosons and fermions point to an underlying {\it supersymmetry} between  
mesons and baryons in QCD.  In fact, the supersymmetric Light Front Holographic approach to QCD not only provides a unified spectroscopy of mesons and baryons, but it also predicts the existence and spectroscopy of tetraquarks:
the mass degeneracy of mesons and baryons with their tetraquark partners, bound states of $3_C$ diquarks and $\bar 3_C$ anti-diquarks.
The meson-baryon-tetraquark 4-plet predicted by the LF supersymmetric approach is illustrated in Fig.~5.  The baryon has two entries in the 4-plet,  analogous to the upper and lower spinor components of a Dirac spinor. 
For example, the  proton  $|[ud] u \rangle $ with $J^z=+1/2$ has equal probability to be a bound state of a scalar $[ud]$  diquark and a $u$ quark with $S^z=+1/2, L^z=0$ or the $u$ quark with nonzero orbital angular momentum $S^z=-1/2, L^z=+1$.  The spin-flip matrix element of the electromagnetic current  between these two states gives the proton's Pauli form factor in the light-front formalism~\cite{Brodsky:1980zm}.

The holographic theory incorporates the dependence on the total quark spin, $S = 0$ for the $\pi$ Regge trajectory, and $S = 1$ for the $\rho$  trajectory, as given by the additional term $2 \kappa^2 S$,  where $S = 0, 1$, in the LF Hamiltonian.  This leads, for example to the correct prediction for the  $\pi - \rho$ mass gap: $M_\rho^2 - M^2_\pi = 2\kappa^2$. In order to describe the quark spin-spin interaction, which distinguishes for example the nucleons from $\Delta$ particles, one includes an identical term, $2 \kappa^2 S$, with $S = 0, 1$ in the LF baryon Hamiltonian which maintains hadronic supersymmetry. The prediction for the mass spectrum of mesons, baryons and tetraquarks is given by~\cite{Brodsky:2016yod}
\begin{align}
M^2_{M \perp} &= 4 \kappa^2 (n+ L_M ) + 2 \kappa^2 S,\\
M^2_{B \perp} &= 4  \kappa^2 (n+ L_B+1) + 2 \kappa^2  S,\\
M^2_{T \perp} & = 4  \kappa^2  (n+ L_T+1) + 2 \kappa^2 S,
\end{align}
with the same slope $\lambda= \kappa^2$ in $L$ and $n$, the radial quantum number.  
The  Regge spectra of the pseudoscalar $S=0$  and vector $S=1$  mesons  are then
predicted correctly, with equal slope in the principal quantum number $n$ and the internal orbital angular momentum.  The nonperturbative pion distribution amplitude 
$\phi_\pi(x) \propto f_\pi \sqrt{x(1-x)}$ predicted by LF holography is  consistent with the Belle data for the photon-to-pion transition form factor~\cite{Brodsky:2011xx}. 
The prediction for the LF wave function $\psi_\rho(x,k_\perp)$ of the  $\rho$ meson gives excellent 
predictions for the observed features of diffractive $\rho$ electroproduction $\gamma^* p \to \rho  p^\prime$~\cite{Forshaw:2012im}.
The prediction for the valence LF wave function of the pion is shown in Fig. \ref{Bledslides6.pdf}.

These predictions for the meson, baryon and tetraquark spectroscopy are specific to zero mass quarks.
In a recent paper~\cite{deTeramond:2021yyi}, we have shown that the breaking of chiral symmetry in holographic light-front QCD from nonzero quark masses is encoded in the longitudinal dynamics, independent of $\zeta$.
The results for $M^2= M^2_\perp + M^2_L$, where $M^2_L$ is the longitudinal contribution from the nonzero quark mass, retains the zero-mass chiral property of the pion predicted by the superconformal algebraic structure which governs its transverse dynamics. The mass scale in the longitudinal light-front Hamiltonian determines the confinement strength in this direction; It is also responsible for most of the light meson ground state mass, consistent with the standard Gell-Mann-Oakes-Renner constraint.    Longitudinal confinement and the breaking of chiral symmetry are found to be different manifestations of the same underlying dynamics that appears in the 't Hooft large-$N_C$ QCD(1 + 1) model.  One also obtains spherical symmetry of the 3-dimensional confinement potential in the nonrelativistic limit.  For related work, see Refs. \cite{Li:2021jqb, Ahmady:2021lsh, Ahmady:2021yzh, Weller:2021wog}.

Phenomenological extensions of the holographic QCD approach  have also led to nontrivial connections between the dynamics of form factors and polarized and unpolarized quark distributions with pre-QCD nonperturbative approaches such as Regge theory and the Veneziano model~\cite{Sufian:2016hwn, deTeramond:2018ecg, Liu:2019vsn}. As discussed in the next section, it also predicts the analytic behavior of the QCD coupling $\alpha_s(Q^2)$ in the nonperturbative domain~\cite{Brodsky:2010ur, Deur:2014qfa}.

\begin{figure}
 \begin{center}
\includegraphics[height=10cm,width=15cm]{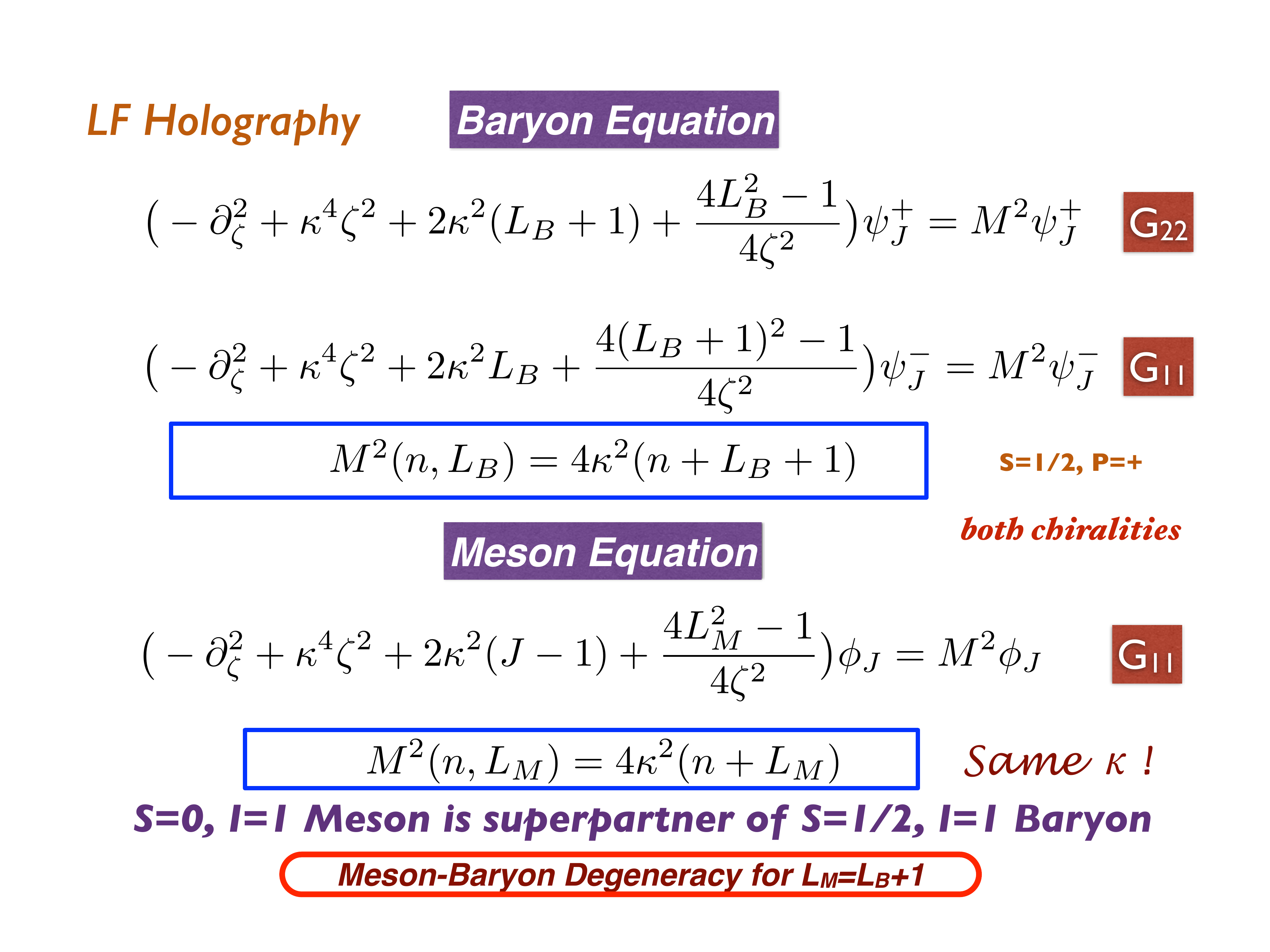}

\end{center}
\caption {The LF Schr\"odinger equations for baryons and mesons for zero quark mass derived from the Pauli $2\times 2$ matrix representation of superconformal algebra.  
The $\psi^\pm$  are the baryon quark-diquark LFWFs where the quark spin $S^z_q=\pm 1/2$ is parallel or antiparallel to the baryon spin $J^z=\pm 1/2$.   The meson and baryon equations are identical if one identifies a meson with internal orbital angular momentum $L_M$ with its superpartner baryon with $L_B = L_M-1.$
See Refs.~\cite{deTeramond:2014asa,Dosch:2015nwa,Dosch:2015bca}.}
\label{FigsJlabProcFig3.pdf}
\end{figure} 
The LF Schr\"odinger Equations for baryons and mesons derived from superconformal algebra  are shown  in Fig. \ref{FigsJlabProcFig3.pdf}.
The comparison between the meson and baryon masses of the $\rho/\omega$ Regge trajectory with the spin-$3/2$ $\Delta$ trajectory is shown in Fig. \ref{FigsJlabProcFig3.pdf}.
Superconformal algebra  predicts the meson and baryon masses are identical if one identifies a meson with internal orbital angular momentum $L_M$ with its superpartner baryon with $L_B = L_M-1.$   Notice that the twist  $\tau = 2+ L_M = 3 + L_B$ of the interpolating operators for the meson and baryon superpartners are the same.   Superconformal algebra also predicts that the LFWFs of the superpartners are identical, and thus they have identical dynamics, such their elastic and transition form factors.   These features can be tested for spacelike  form factors at  JLab12.

The extension of light-front QCD  to superconformal algebra has leads to a specific mass degeneracy between mesons, baryons and tetraquarks~\cite{deTeramond:2014asa, Dosch:2015nwa, Brodsky:2016yod}  underlying the $SU(3)_C$ representation properties, since a diquark cluster has the same color-triplet representation as an antiquark, namely $\bar 3 \in 3 \times 3$.  The meson wave function $\phi_M$, the upper and lower components of the baryon wave function, $\phi_{B \,\pm}$, and the tetraquark wave function, $\phi_T$, can be arranged as a supersymmetric 4-plet matrix~\cite{Brodsky:2016yod, Zou:2018eam}  
\begin{align}
\vert \Phi \rangle =   \begin{pmatrix}
    \phi_M^{\, (L+1)} & \phi_{B \, -}^{\, (L + 1)}\\
    \phi_{B \, + } ^{\, (L)} & \phi_T^{\, (L)}
    \end{pmatrix} ,
\end{align}
with $H^{LF} \vert \Phi \rangle = M^2 \vert \Phi \rangle$ and $L_M = L_B + 1$,  $L_T = L_B$. The constraints from superconformal structure uniquely determine the form of the effective transverse confining potential for mesons, nucleons and tetraquarks~\cite{deTeramond:2014asa, Dosch:2015nwa, Brodsky:2016yod}, 
and lead to the remarkable relations  $L_M = L_B + 1$, $L_T = L_B$. The superconformal algebra also predicts the universality of Regge slopes 
with a unique scale $\lambda=\kappa^2$ for all hadron families.

\section {The QCD Coupling at all Scales} 
\begin{figure}
\begin{center}
\includegraphics[height=7cm,width=12cm]{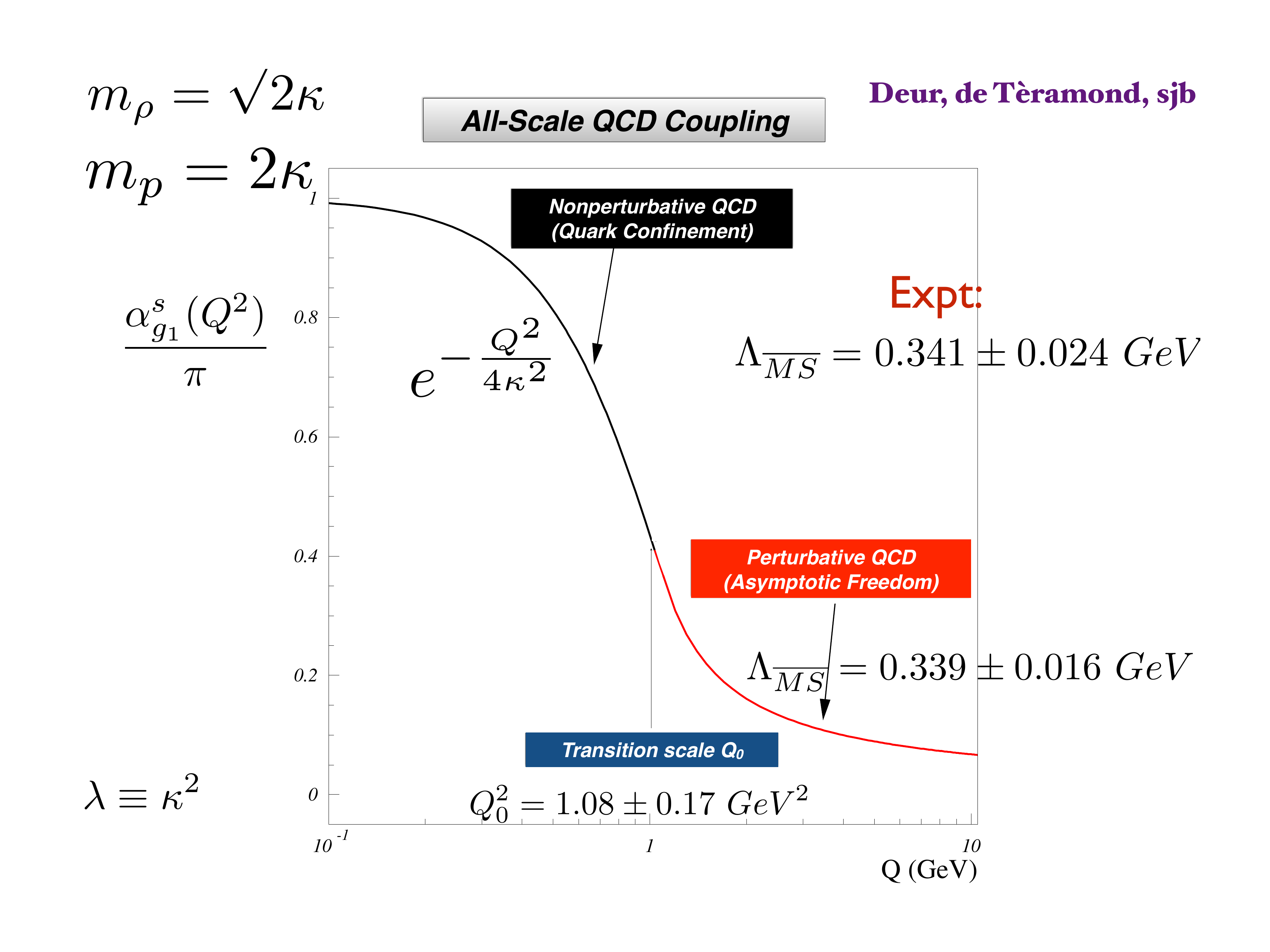}
\includegraphics[height=7cm,width=12cm]{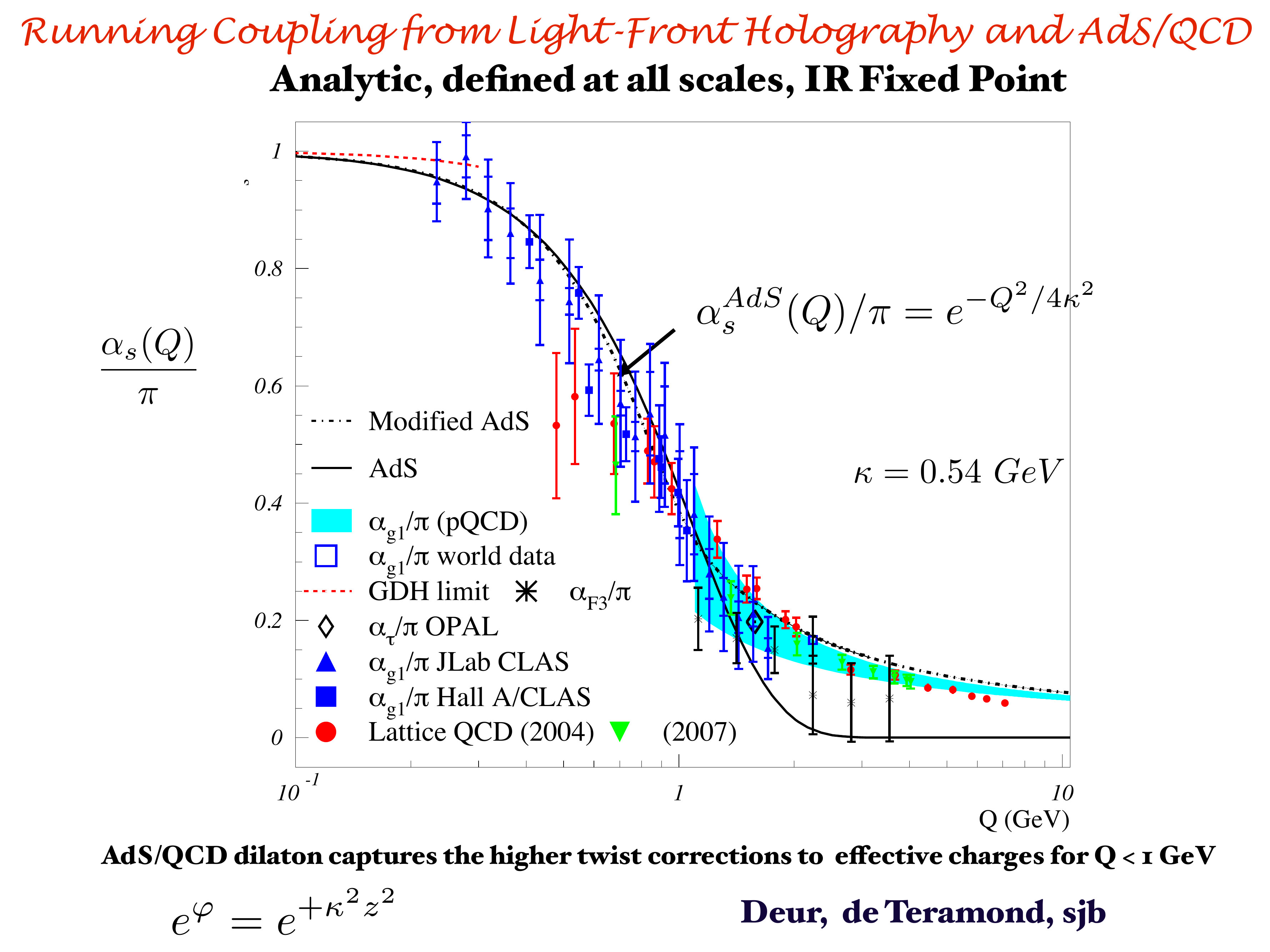}
\end{center}
\caption{
(A). Prediction from LF Holography for the QCD running Coupling $\alpha^s_{g_1}(Q^2)$.   The magnitude and derivative of the perturbative and nonperturbative coupling are matched at the scale $Q_0$.  This matching connects the perturbative scale 
$\Lambda_{\overline{MS}}$ to the nonperturbative scale $\kappa$ which underlies the hadron mass scale. 
(B). Comparison of the predicted nonperturbative coupling with measurements of the effective charge $\alpha^s_{g_1}(Q^2)$  
defined from the Bjorken sum rule.  
See Ref.~\cite{Brodsky:2014jia}. 
}
\label{FigsJlabProcFig5.pdf}
\end{figure}

The QCD running coupling can be defined~\cite{Grunberg:1980ja} at all momentum scales from any perturbatively calculable observable, such as the coupling $\alpha^s_{g_1}(Q^2)$ which is defined from measurements of the Bjorken sum rule.   At high momentum transfer, such ``effective charges"  satisfy asymptotic freedom, obey the usual pQCD renormalization group equations, and can be related to each other without scale ambiguity 
by commensurate scale relations~\cite{Brodsky:1994eh}.  
The dilaton  $e^{+\kappa^2 z^2}$ soft-wall modification~\cite{Karch:2006pv} of the AdS$_5$ metric, together with LF holography, predicts the functional behavior 
in the small $Q^2$ domain~\cite{Brodsky:2010ur}: 
${\alpha^s_{g_1}(Q^2) = 
\pi   e^{- Q^2 /4 \kappa^2 }}. $ 
Measurements of  $\alpha^s_{g_1}(Q^2)$ are remarkably consistent with this predicted Gaussian form. 
We have also shown how the parameter $\kappa$,  which   determines the mass scale of  hadrons in the chiral limit, can be connected to the  mass scale $\Lambda_s$  controlling the evolution of the perturbative QCD coupling~\cite{Brodsky:2010ur,Deur:2014qfa,Brodsky:2014jia}. This connection can be done for any choice of renormalization scheme, including the $\overline{MS}$ scheme,
as seen in  Fig.~\ref{FigsJlabProcFig5.pdf}. 
The relation between scales is obtained by matching at a scale $Q^2_0$ the nonperturbative behavior of the effective QCD coupling, as determined from light-front holography, to the perturbative QCD coupling with asymptotic freedom.
The result of this perturbative/nonperturbative matching is an effective QCD coupling  which is defined at all momenta.

\section{Light-Front Wave Functions and QCD}

Measurements of hadron structure -- such as the structure functions determined by  deep inelastic lepton-proton scattering (DIS) -- are analogous to a flash photograph: one observes the hadron at fixed 
$\tau=  t+z/c$ along a light-front, not at a given instant of time $t$.  The underlying physics follows from the 
the light-front  wave functions (LFWFs)  
$\psi_n(x_i,  \vec k_{\perp i },  \lambda_i )$ with 
$x_i = {k^+_i\over P^+} = {k^0_i+k^z_i\over P^0+P^z}, \sum^n_i x_1 =1, \sum^n_i \vec k_{\perp _i} =\vec 0_\perp$  and spin projections $\lambda_i$.  The LFWFs are the Fock state projections of the eigenstates of the QCD LF invariant Hamiltonian $H_{LF} |\Psi\rangle = M^2|\Psi\rangle$~\cite{Brodsky:1997de}, where the LF Hamiltonian is the light-front time evolution operator defined directly from the QCD Lagrangian. 
One can avoid ghosts and longitudinal  gluonic degrees of freedom by choosing to work in the light-cone gauge  $A^+ =0$.  
The LFWFs are boost invariant; i.e., they are independent of the hadron's momentum $P^+ =P^0 +P^z, \vec P_\perp.$
This contrasts with the wave functions defined at a fixed time $t$ -- the Lorentz boost of an instant-form wave function is much more complicated than a 
Melosh transform~\cite{Brodsky:1968ea} -- even the number of Fock constituents changes under a boost.  
Current matrix elements such as form factors are simple overlaps of  the initial-state and final-state LFWFs, as given by the 
Drell-Yan-West formula~\cite{Drell:1969km, West:1970av, Brodsky:1980zm}. There is no analogous formula for the instant form, since one must take into account the coupling of the external current to connected vacuum-induced currents.
Observables such as structure functions, transverse momentum distributions, and distribution amplitudes are defined from the hadronic LFWFs.  
The distribution amplitudes $\phi_H(x_i, Q) $ are given by the valence LFWF integrated over transverse momentum $k^2_\perp < Q^2$. 

Since they are frame-independent, the structure functions measured in DIS are  the same whether they are measured in an electron-proton collider or in a fixed-target experiment where the proton is at rest.     There is no concept of length contraction of the hadron or nucleus at  a collider -- no collisions of  ``pancakes" --   since the observations  of the collisions of the composite hadrons are made at fixed $\tau$, not  at fixed time.    The dynamics of a hadron in the LF formalism is not dependent on the observer's Lorentz frame. 
Hadron form factors are matrix elements of the noninteracting electromagnetic current $j^\mu$  of  the hadron, as in the interaction picture of  quantum mechanics. One chooses the frame where the virtual photon 4-momentum $q^\mu$ has $q^+=0$, ${\vec q_\perp}^2=Q^2 = -q^2$ and $q^- P^+ = q\cdot p.$  One can also choose to evaluate matrix elements of $j^+= j^0+ j^z$ which eliminates matrix elements between Fock states with and extra $q \bar q$ pair.

The frame-independent LF Heisenberg equation $H^{QCD}_{LF} |\psi_H\rangle = M^2_H \psi_H\rangle $ can be solved numerically by matrix diagonalization of the LF Hamiltonian in LF Fock space  using ``Discretized Light-Cone  Quantization" (DLCQ)~\cite{Pauli:1985pv}, where anti-periodic boundary conditions in 
$x^-$ render the $k^+$ momenta  discrete  as well as  limiting the size of the Fock basis.  In fact, one can easily solve 1+1 quantum field theories such as QCD$(1+1)$ ~\cite{Hornbostel:1988fb} for any number of colors, flavors, and quark masses.  Unlike lattice gauge theory, the nonperturbative DLCQ analysis is in Minkowski space, it is frame-independent, and it is free of fermion-doubling problems.    A new method for solving nonperturbative QCD ``Basis Light-Front Quantization" (BLFQ)~\cite{Vary:2009gt,Vary:2014tqa},  uses the eigensolutions of a color-confining approximation to QCD (such as LF holography ) as the basis functions,  rather than the plane-wave basis used in DLCQ.  The LFWFs can also be determined from covariant Bethe-Salpeter wave function by 
integrating over $k^-$~\cite{Brodsky:2015aia}.  In fact, advanced quantum computers are now being used to obtain the DLCQ and BLFQ solutions.

Factorization theorems as well as the DGLAP and ERBL evolution equations for structure functions and distribution amplitudes, respectively, can be derived using the light-front Hamiltonian formalism~\cite{Lepage:1980fj}.  In the case of an electron-ion collider, one can represent the cross section for $e - p $ collisions as a convolution of the hadron and virtual photon structure functions times the subprocess cross-section in analogy to hadron-hadron collisions.   This description of $\gamma^* p \to X$ reactions  gives new insights into electroproduction physics such as the dynamics of heavy quark-pair production, where 
intrinsic heavy quarks play an important role~\cite{Brodsky:2015uwa}.

In the case of $e p \to e^\prime X$, one can consider the collisions of the confining  QCD flux tube appearing between the $q$ and $\bar q$  of the virtual photon with the flux tube between the quark and diquark of the proton.   Since the $q \bar q$ plane is aligned with the scattered electron's plane, the resulting ``ridge"  of hadronic multiplicity produced from the $\gamma^* p$ collision will also be aligned with the scattering plane of the scattered electron.  The virtual photon's flux tube will also depend on the photon virtuality $Q^2$, as well as the flavor of the produced pair arising from $\gamma^* \to q \bar q$.  The resulting dynamics~\cite{Brodsky:2014hia} is  a natural extension of the flux-tube collision description of the ridge produced in $p-p$ collisions~\cite{Bjorken:2013boa}.

\section{Other Features of Light-Front QCD}

There are a number of advantages if one uses  LF Hamiltonian methods for perturbative QCD calculations.  The LF formalism is frame-independent and causa.  If one chooses LF gauge $A^+=0$ the gluons have only transverse polarization and no ghosts. If one chooses the frame $q^+=0$ the current does not create pairs.  Unlike instant form, where one must sum  over $n !$ frame-dependent  amplitudes, only the $\tau$-ordered diagrams where every line has  positive $k^+ =k^0+k^z$  can contribute~\cite{Cruz-Santiago:2015dla}. The number of nonzero amplitudes is also greatly reduced by noting that the total angular momentum projection $J^z = \sum_i^{n-1 } L^z_i + \sum^n_i S^z_i$ and the total $P^+$ are  conserved at each vertex.  In addition, in a renormalizable theory the change in orbital angular momentum is limited to $\Delta L^z =0,\pm 1$ at each vertex.  The calculation of a subgraph of any order in pQCD only needs to be done once;  the result can be stored in a ``history" file, since in light-front perturbation theory,  the numerator algebra is independent of the process; the denominator changes, but only by a simple shift of the initial $P^-$. Loop integrations are three-dimensional: $\int d^2\vec k_\perp \int^1_0 dx.$
Renormalization can be done using the ``alternate denominator" method which defines the required subtraction counter-terms~\cite{Brodsky:1973kb}.

The LF vacuum in LF Hamiltonian theory is defined as the eigenstate of $H_{LF}$ with lowest invariant mass. Since propagation of particles with negative $k^+$  does not appear, there are no loop amplitudes appearing in the LF vacuum -- it is  is thus trivial up to possible $k^+=0$ ``zero"  modes.   The usual quark and gluon QCD vacuum condensates of the instant form =are replaced by physical effects,  such as the running quark mass and the physics contained within the hadronic LFWFs  in the hadronic domain. This is referred to as ``in-hadron" condensates~\cite{Casher:1974xd,Brodsky:2009zd,Brodsky:2010xf}.  In the case of the Higgs theory, the traditional Higgs vacuum expectation value (VEV) is replaced by a zero mode, analogous to a classical 
Stark or Zeeman field.~\cite{Srivastava:2002mw}   This approach contrasts with the traditional view of the vacuum  based on the instant form. 

The instant-form vacuum, the lowest energy eigenstate of the instant-form Hamiltonian,  is defined at one instant of time over all space; it is thus acausal and frame-dependent.  It is usually argued that the QCD contribution to the cosmological constant -- dark energy  -- is $10^{45}$ times larger that observed, and in the case of the Higgs theory, the Higgs VEV is argued to be $10^{54}$ larger than observed~\cite{Zee:2008zz}, 
estimates based on the loop diagrams of the acausal frame-dependent instant-form vacuum.  However, the universe is observed within the causal horizon, not at a single instant of time.  In contrast, the light-front vacuum provides a viable description of the visible universe~\cite{Brodsky:2010xf}. Thus, in agreement with Einstein's theory of general relativity, quantum effects do not contribute to the cosmological constant.   In the case of the Higgs theory, the Higgs zero mode has no energy density,  so again  it gives no contribution to the cosmological constant.  However, it is possible that if one solves the Higgs theory in a curved universe, the zero mode will be replaced with a field of nonzero curvature which could give a nonzero contribution.

\section{Gluon matter distribution in the proton and pion from extended holographic light-front QCD}
The holographic light-front QCD framework provides a unified nonperturbative description of the hadron mass spectrum, form factors and quark distributions.
In a recent article ~\cite{deTeramond:2021lxc}  we have extended our previous description of quark distributions \cite{deTeramond:2018ecg,Liu:2019vsn} 
in LF holographic QCD to predict 
the gluonic distributions of both the proton and pion from the coupling of the metric fluctuations induced by the spin-two Pomeron with the energy momentum tensor in anti-de Sitter space, together with constraints imposed by the Veneziano model without additional free parameters. The gluonic and quark distributions are shown to have significantly different effective QCD mass scales.
The comparison of our predictions with the gluon gravitational form factor computed from Euclidean lattice gauge theory and the gluon distribution in the proton and pion from global analyses also give very good results.

\section{Intrinsic Heavy Quarks}

Quantum Chromodynamics (QCD), the underlying theory of strong interactions, with quarks and gluons as the fundamental degrees of freedom, predicts that the heavy quarks in the nucleon-sea to have both perturbative ``extrinsic" and nonperturbative ``intrinsic" origins.  The extrinsic sea arises from gluon splitting which is triggered by a probe in the reaction. It can be calculated order-by-order in perturbation theory.  In contrast, the intrinsic sea is encoded in the nonperturbative wave functions of the nucleon eigenstate. 

The existence of nonperturbative intrinsic charm (IC) was originally proposed in the BHPS model~\cite{Brodsky:1980pb} and developed further in subsequent papers~\cite{Brodsky:1984nx,Harris:1995jx,Franz:2000ee}. The intrinsic contribution to the heavy quark  distributions of hadrons at high $x$ corresponds to Fock states such as  $|uud Q \bar Q\rangle$ where the heavy quark 
pair is multiply connected to two or more valence quarks of the proton, in distinction to the higher order corrections to DGLAP evolution. The LF wave function is maximal at minimal off-shellness; i.e., when the constituents all have the same rapidity  $y_i$, and thus 
$x_i \propto \sqrt{(m_i^2+ { \vec k_{\perp i}}^2 )}$.  Here $x= {k^+\over P^+} = {k^0 + k^3\over P^0 + P^3}$ is the frame-independent light-front momentum fraction carried by the heavy quark in a hadron with momentum $P^\mu$. 
In the case of deep inelastic lepton-proton scattering, the LF momentum fraction variable $x$  in the proton structure functions can be identified with the Bjorken variable 
$x = {Q^2\over 2 p \cdot q}.$
These heavy quark contributions 
to the nucleon's PDF thus peak at large $x_{bj}$ and thus have important  implication for LHC and EIC collider phenomenology, including Higgs and heavy hadron production at high $x_F$~\cite{Royon:2015eya}.
It also opens up new opportunities to study heavy quark phenomena in fixed target experiments such as the proposed AFTER~\cite{Brodsky:2015fna} fixed target facility at CERN.  Other applications are presented in Refs.~\cite{Brodsky:2020zdq,Bednyakov:2017vck,Brodsky:2016fyh}.
The  existence of intrinsic heavy quarks also illuminates fundamental aspects of nonperturbative QCD.

In  Light-Front Hamiltonian theory, the intrinsic heavy quarks of the proton are associated with non-valence Fock states. 
such as $|uud Q \bar Q \rangle$ in the hadronic eigenstate of the LF Hamiltonian; this implies that the heavy quarks are multi-connected to the valence quarks. The probability for the heavy-quark Fock states scales as $1/m^2_Q$ in non-Abelian QCD.  Since the LF wave function is maximal at minimum off-shell invariant mass; i.e., at equal rapidity, the intrinsic heavy quarks carry large momentum fraction $x_Q$.  A key characteristic is different momentum and spin distributions for the intrinsic $Q$ and $\bar Q$ in the nucleon; for example the charm-anticharm asymmetry, since the comoving quarks are sensitive to the global quantum numbers of the nucleon~\cite{Brodsky:2015fna}.  Furthermore, since all of the  intrinsic quarks in the $|uud Q \bar Q\rangle$  Fock state have similar rapidities as the valence quarks, they can re-interact, leading to significant $Q$ vs $\bar Q$ asymmetries.  The concept of intrinsic heavy quarks was also proposed in the context of  meson-baryon fluctuation models~\cite{Pumplin:2005yf, Navarra:1995rq}, where intrinsic charm was identified with two-body state $\bar{D}^0(u\bar{c})\Lambda^+_c(udc)$ in the proton. This identification  predicts large asymmetries in the charm versus anti-charm momentum and spin distributions,  Since these heavy quark distributions depend on the correlations determined by the valence quark distributions, they are referred to as {\it  intrinsic } contributions to the hadron's fundamental structure. A specific analysis of the intrinsic charm content of the deuteron is given in Ref.~\cite{Brodsky:2018zdh}.
In contrast, the contribution to the heavy quark PDFs arising from gluon splitting are symmetric in $Q$ vs $\bar Q$. The contributions generated by DGLAP evolution at low $x$ can be considered as  {\it extrinsic} contributions since they only depend on the gluon distribution. The gluon splitting contribution to the heavy-quark degrees of freedom is  perturbatively calculable 
using  DGLAP
evolution.  To first approximation, the perturbative extrinsic heavy quark distribution falls as $(1-x)$ times the gluon distribution and is limited to low $x_{bj}.$
Thus, unlike the conventional $\log m^2_Q$ dependence of the low $x$  extrinsic gluon-splitting contributions, the probabilities for the intrinsic heavy quark Fock states at high $x$  scale as $1\over m_Q^2$  in non-Abelian QCD, and the relative probability of intrinsic bottom to charm is of order ${m^2_c\over m^2_b} \sim  {1\over 10}.$
In contrast, the probability for a higher Fock state containing heavy leptons in a QED atom  scales as $1\over m_\ell^4$, corresponding to the twist-8 Euler-Heisenberg light-by-light self-energy insertion.  Detailed derivations based on the OPE have been given in Ref.~\cite{Brodsky:1984nx,Franz:2000ee}.

%%%%%%%
%%%%%%%
\begin{figure}[htp]
\begin{center}
\setlength\belowcaptionskip{-2pt}
\includegraphics[width=3.2in, height=2.3in]{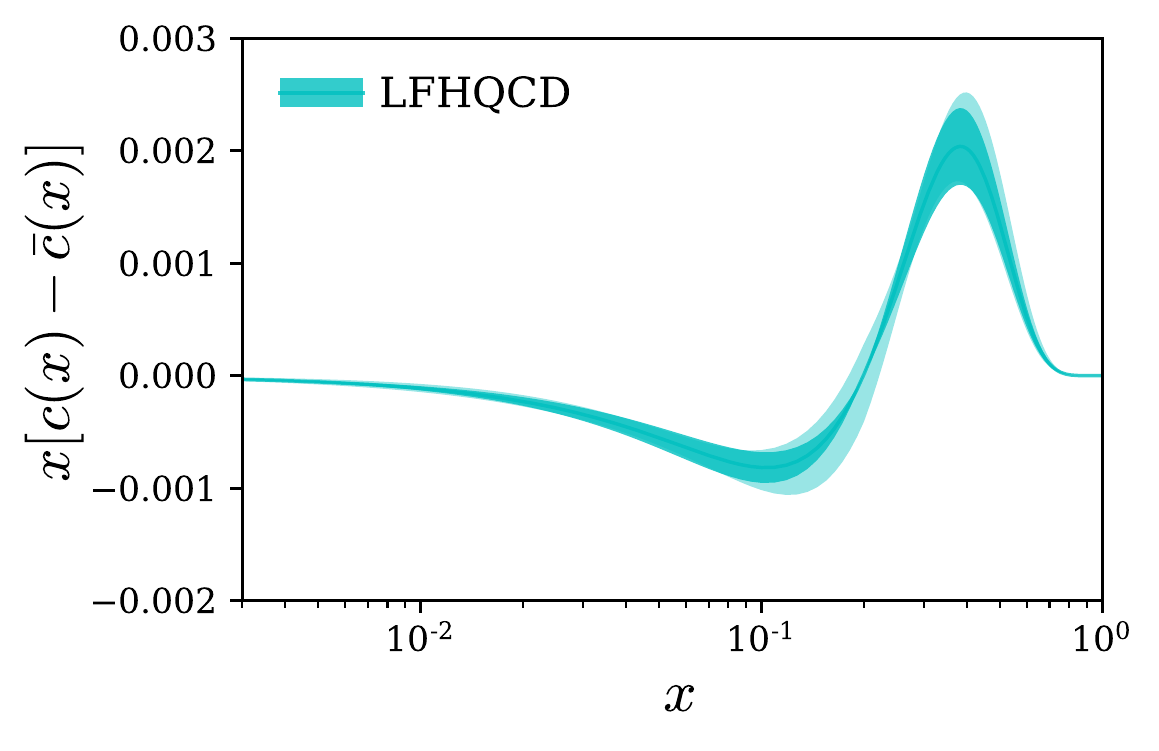}
\caption{The difference of charm and anticharm structure functions $x[c(x)-\bar{c}(x)]$ obtained from the LFHQCD formalism using the lattice QCD input of charm electromagnetic form factors $G^c_{E,M}(Q^2)$. \label{fig:ccbardis}  
The outer cyan band indicates an estimate of systematic uncertainty in the $x[c(x)-\bar{c}(x)]$ distribution obtained from a variation of  the hadron scale $\kappa_c$ by 5\%.  From Ref.~\cite{Sufian:2020coz}.}
\end{center}
\end{figure}
%%%%%%%
%%%%%%%

In an important recent development~\cite{Sufian:2020coz},  the difference of the charm and anticharm  quark distributions in the proton, $\Delta c(x) = c(x) -\bar c(x)$,  has been computed from first principles in QCD using lattice gauge theory.  A key  theoretical tool is the computation of the charm and anticharm quark contribution 
to the electromagnetic form factor of the proton which would vanish if $c(x) =\bar c(x).$    The exclusive-inclusive connection, together with the LFHQCD formalism, predicts the asymmetry of structure functions $c(x)- \bar c(x)$ which is also odd under charm-anticharm interchange.   
The predicted $c(x)- \bar c(x)$ distribution is large and nonzero at large at $x \sim 0.4$, consistent with the expectations of intrinsic charm. See Fig.~\ref{fig:ccbardis}.  

The $c(x)$ vs. $\bar c(x)$  asymmetry can also be understood physically by identifying the $ |uud c\bar c \rangle$ Fock state with the $|\Lambda_{udc} D_{u\bar c} \rangle$ off-shell excitation of the proton.

 A related application of lattice gauge theory to the nonperturbative strange-quark sea from lattice QCD is given in Ref.~\cite{Sufian:2018cpj}.

\begin{figure}
 \begin{center}
\includegraphics[height= 10cm,width=15cm]{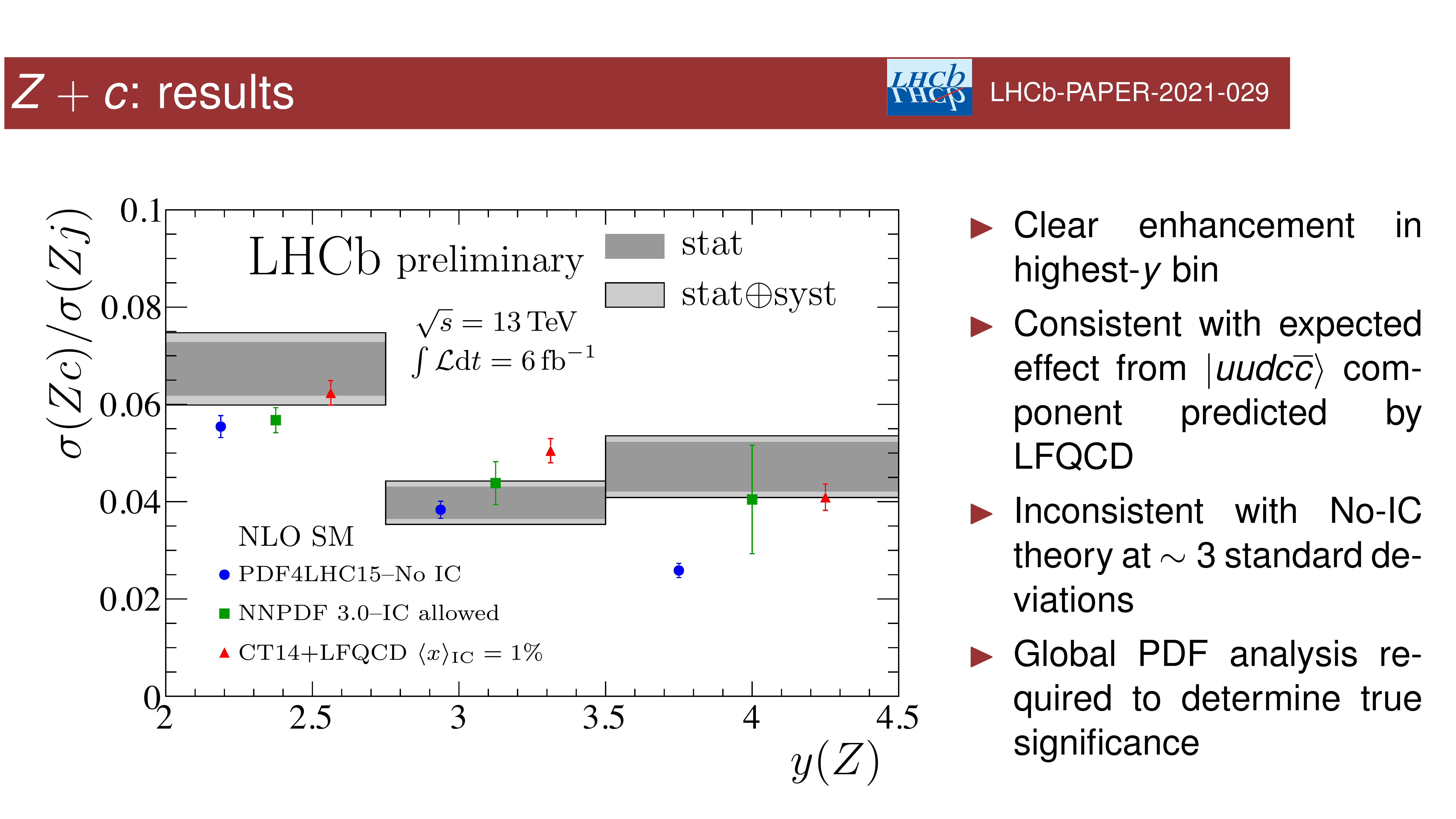}
\end{center}
\caption{The charm distribution in the proton determined from LHCb  measurements of 
$Z$ bosons produced in association with charm at forward rapidity~\cite{LHCb:2021stx}.}
\label{Bledslides1LHCb.pdf}
\end{figure}

There have been many phenomenological calculations involving the existence of a non-zero IC  component which can explain anomalies in the experimental data and to predict  its novel signatures of IC in upcoming experiments~\cite{Brodsky:2015fna}.   A recent measurement by LHCb is shown in Fig. 10.
The observed spectrum exhibits a sizable enhancement at forward Z rapidities, consistent with the effect expected if the proton contains the $ |uud \bar c c \rangle$ Fock state predicted by LFQCD.~\cite{LHCb:2021stx}

Thus QCD predicts two separate and distinct contributions to the heavy quark distributions $q(x,Q^2)$ of  the nucleons at low and high $x$.
Here $x= {k^+\over P^+} = {k^0 + k^3\over P^0 + P^3}$ is the frame-independent light-front momentum fraction carried by the heavy quark in a hadron with momentum $P^\mu$. 
In the case of deep inelastic lepton-proton scattering, the LF momentum fraction variable $x$  in the proton structure functions can be identified with the Bjorken variable 
$x = {Q^2\over 2 p \cdot q}.$
At small $x$,  heavy-quark pairs are dominantly produced via the standard gluon-splitting subprocess $g \to Q \bar Q$.

\section{Color Transparency~\cite{sjbGdT} }

One of the most striking properties of QCD phenomenology is ``color transparency"~\cite{Brodsky:1988xz}, the reduced absorption of a hadron as it propagates through nuclear matter, if it is produced at high transverse momentum in a hard exclusive process, such as elastic lepton-proton scattering. The nuclear absorption reflects the size of the color dipole moment of the propagating hadron; i.e.,  the separation between its colored constituents.

The key quantity which measures the  transverse size of a scattered hadron in a given Fock state is~\cite{sjbGdT}
$a_\perp = \sum_{i=1}^{n-1} x_i b_{\perp i}$. 
The LF QCD formula for form factors can then be written compactly in impact space as 
\begin{align}
F(Q^2) = \int^1_0 dx d^2 a_\perp  e^{i \vec q_\perp\cdot a_\perp} q(x, a_\perp),
\end{align}
and thus 
$\langle a^2_\perp(Q^2)\rangle  = - 4{{d \over dQ^2} F(Q^2) \over F(Q^2)}$
measures the slope of the hadron factor. 
We can use LF holography to show that $\langle a^2_\perp(Q^2) \rangle_\tau = 4 {\tau-1\over Q^2}$ for a Fock state of twist $\tau$ at large $Q^2$;
thus, as expected, the hadronic size decreases with increasing momentum transfer $Q^2$, and that the size of the hadron increases with its twist $\tau$.

\begin{figure}
 \begin{center}
\includegraphics[height= 10cm,width=15cm]{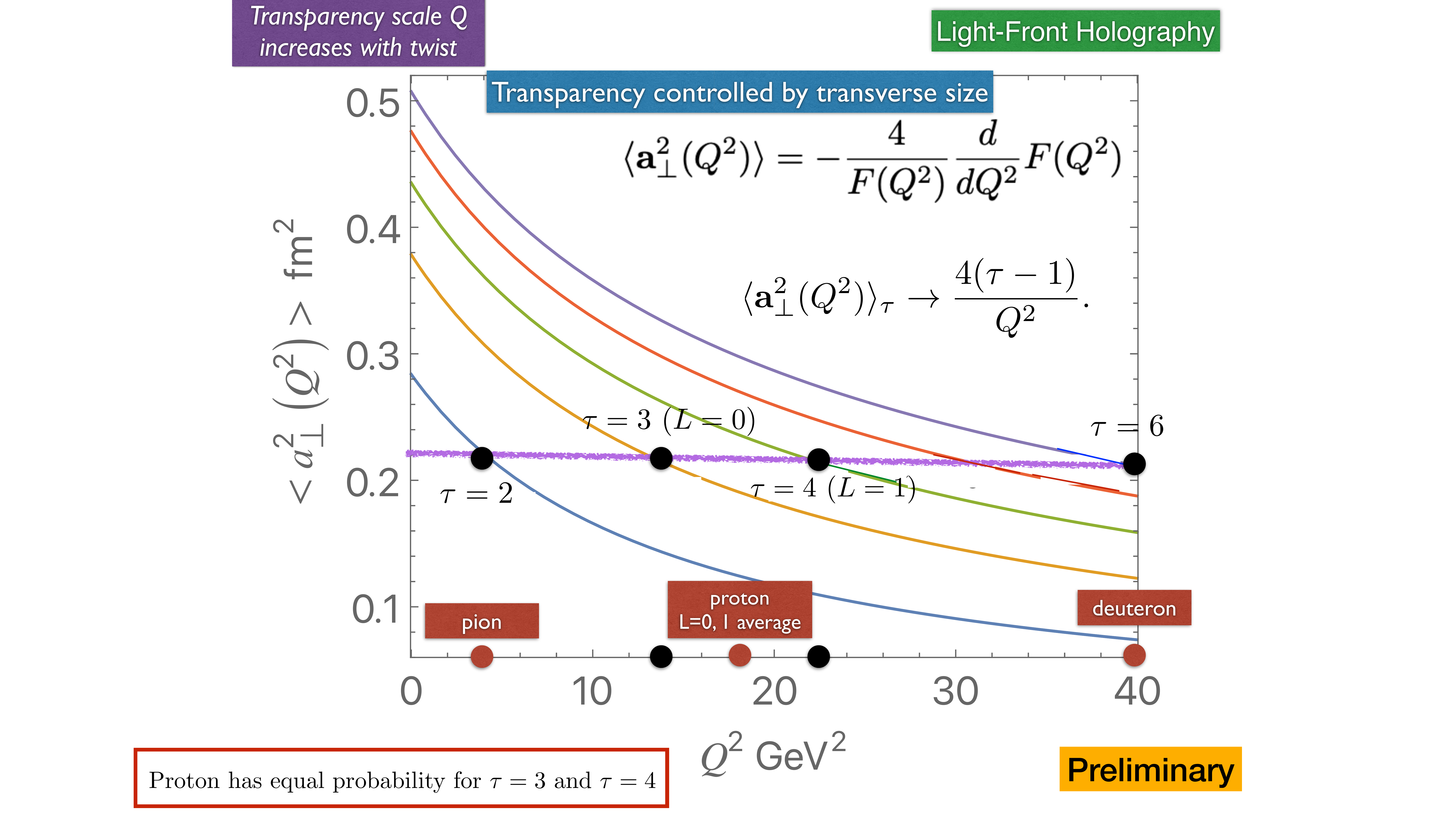}
\end{center}
\caption{Predictions from LF holography for the effective transverse size of hadrons. }
\label{Bledslides-CT.pdf}
\end{figure}

A key prediction is that the size of $a_\perp$ is smaller for mesons $(\tau =2$  than for baryons with $\tau=3,4$, corresponding to the quark-diquark Fock states with $ L=0 $ and $ L=1 $ respectively.
In fact, the proton is predicted to have ``two-stage" color transparency $Q^2>14~GeV^2$ for the $ |[ud] u \rangle$  twist-3 Fock state with orbital angular momentum $ L=0$ 
 and $Q^2 > 16~GeV^2$ for the later onset of CT for its $L=1$ twist-4 component.   See fig. \ref{Bledslides-CT.pdf}
 Note that LF holography predicts equal  quark probability for the $L=0$  and $ L=1$ Fock states.
 Color transparency is thus predicted to occur at a significantly  higher $Q^2$ for baryons  $(Q^2 > 14~GeV^2)$, than for mesons $(Q^2 > 4~GeV^2)$.
 This is consistent with a recent test of color transparency at JLab which has confirmed color transparency for the the $\pi$ and $\rho$ ~\cite{HallC:2020ijh}; however, the measurements
 in this experiment are limited to values below the range of $Q^2$ where proton color transparency is predicted to occur.
 
 \begin{figure}
 \begin{center}
\includegraphics[height= 10cm,width=15cm]{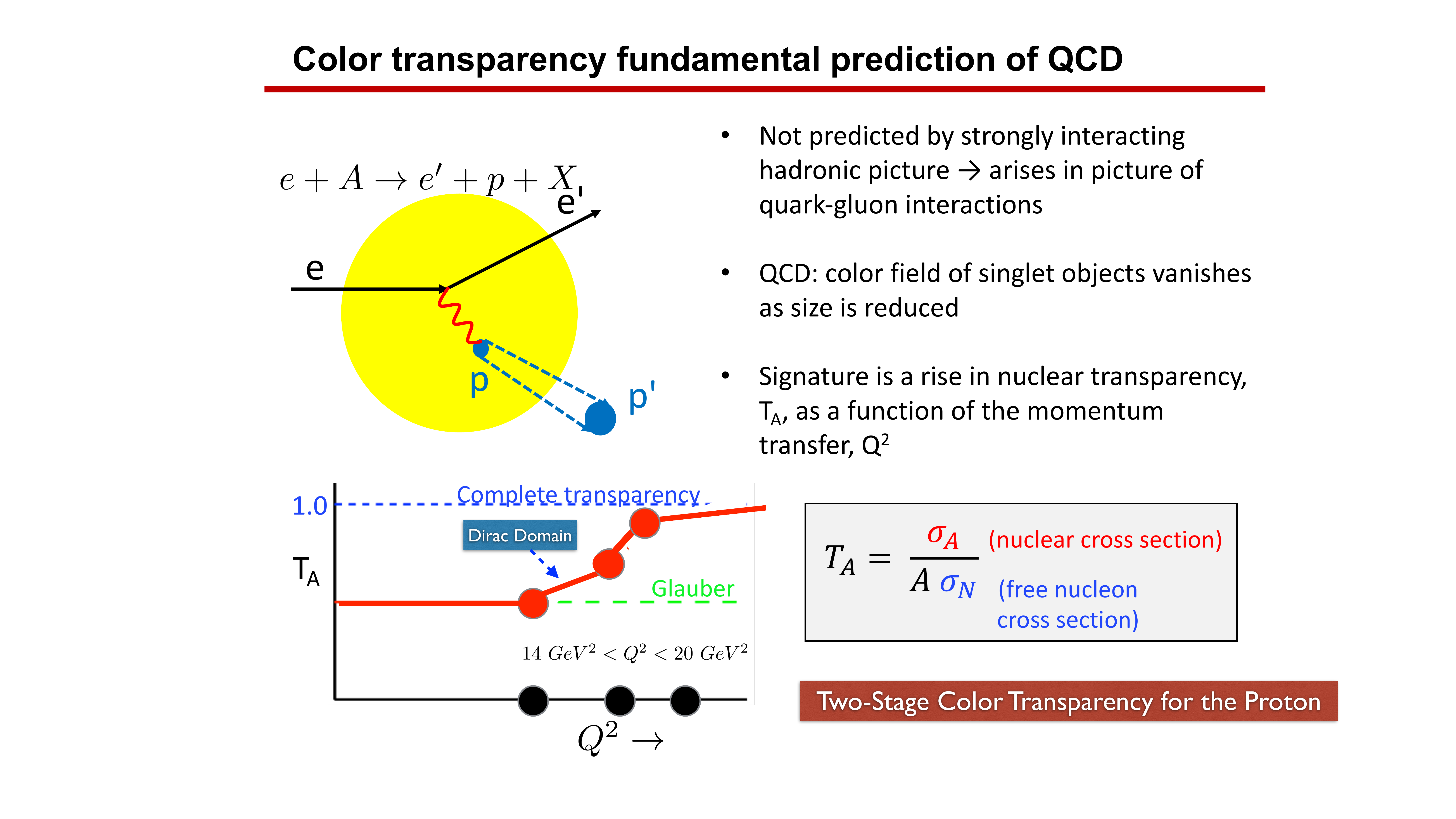}
\end{center}
\caption{Two-stage color transparency and transmission probability of the proton in a nuclear medium from LF Holography. }
\label{Bledslides9.pdf}
\end{figure}

Remarkably, color transparency for the production of an intact deuteron nucleus in $e A \to d + X_{(A-2)}$  quasi-exclusive reactions should be observed at $Q^2 > 50~GeV^2$. This can be tested in $e d \to e d $ elastic scattering in a nuclear background.

It has been speculated~\cite{Caplow-Munro:2021xwi} that the ``Feynman mechanism", where the behavior of the struck quark at $x \sim 1$ in the proton LFWF plays a key role for hard exclusive hadronic processes
does not predict color transparency.  However, LF wave functions are functions of the invariant mass 
$\sum_i {\vec k^2_{\perp i }+ m^2_i \over x_i}$
so that their behavior at large $k_\perp$ and large $x$ are correlated.  Thus color transparency occurs for scattering amplitudes involving both the large transverse momentum and large $x$ domains. The three-dimensional LF spatial symmetry of LFWFs also leads to the 
exclusive-inclusive connection, relating the counting rules for the behavior of form factors at large $Q^2$ and structure functions at $x_{bj} \to 1$.

\section {Removing Renormalization Scale Ambiguities}

It has become conventional to simply guess the renormalization scale and choose an arbitrary range of uncertainty when making perturbative QCD (pQCD) predictions. However, this {\it ad hoc} assignment of the renormalization scale and the estimate of the size of the resulting uncertainty leads to anomalous renormalization scheme-and-scale dependences. In fact, relations between physical observables must be independent of the theorist's choice of the renormalization scheme, and the renormalization scale in any given scheme at any given order of pQCD is not ambiguous. The {\it Principle of Maximum Conformality} (PMC)~\cite{Brodsky:2011ig}, which generalizes the conventional Gell-Mann-Low method for scale-setting in perturbative QED to non-Abelian QCD, provides a rigorous method for achieving unambiguous scheme-independent, fixed-order predictions for observables consistent with the principles of the renormalization group.   The renormalization scale of the running coupling depends dynamically
on the virtuality of the underlying quark and gluon subprocess and thus the specific kinematics
of each event. 

The  key problem in making precise perturbative QCD predictions is
the uncertainty in determining the renormalization scale $\mu$ of
the running coupling $\alpha_s(\mu^2).$ 
The purpose of the running
coupling in any gauge theory is to sum all terms involving the
$\beta$ function; in fact, when the renormalization scale is set
properly, all non-conformal $\beta \ne 0$ terms  in a perturbative
expansion arising from renormalization are summed into the running
coupling. The remaining terms in the perturbative series are then
identical to that of a  conformal theory; i.e., the corresponding
theory with $\beta=0$. 

The renormalization scale in the PMC is fixed such that all $\beta$ nonconformal
terms are eliminated from the perturbative series and are resummed into the running coupling; this procedure results in a convergent, scheme-independent conformal series without factorial renormalon divergences. The resulting scale-fixed predictions for physical observables using
the PMC are also {\it  independent of
the choice of renormalization scheme} --  a key requirement of 
renormalization group invariance.  The PMC predictions are also independent of the choice of the {\it initial} renormalization scale $\mu_0.$  
The PMC thus sums all of the non-conformal terms associated with the QCD $\beta$ function, thus providing a rigorous method for eliminating renormalization scale ambiguities in quantum field theory. 
Other important properties of the PMC are that the resulting series are free of renormalon
resummation problems,  and  the predictions agree with QED scale-setting in the Abelian limit.
The PMC is also the theoretical principle underlying the BLM
procedure, commensurate scale relations between observables, and
the scale-setting method used in lattice gauge theory.  The number
of active flavors $n_f$ in the QCD $\beta$ function is also
correctly determined.   
We have also showed that a single global
PMC scale, valid at leading order, can be derived from basic
properties of the perturbative QCD cross section.   We have given a detailed comparison of these PMC approaches by comparing their predictions for three important quantities 
$R_{e+ e}$, $R_\tau$ and $\Gamma_{H \to b \bar b}$ up to four-loop pQCD corrections~\cite{Brodsky:2011ig}.  The numerical results show that the single-scale PMCs method, which involves a somewhat simpler analysis, can serve as a reliable substitute for the full multi-scale PMCm method, and that it leads to more precise pQCD predictions with less residual scale dependence.
The PMC thus greatly improves the reliability and precision of QCD predictions at the LHC and other colliders~\cite{Brodsky:2011ig}.
 As we have demonstrated, the PMC also has the potential to greatly increase the sensitivity of experiments at the LHC to new physics beyond the Standard Model.

  \begin{figure}
 \begin{center}
\includegraphics[height= 10cm,width=15cm]{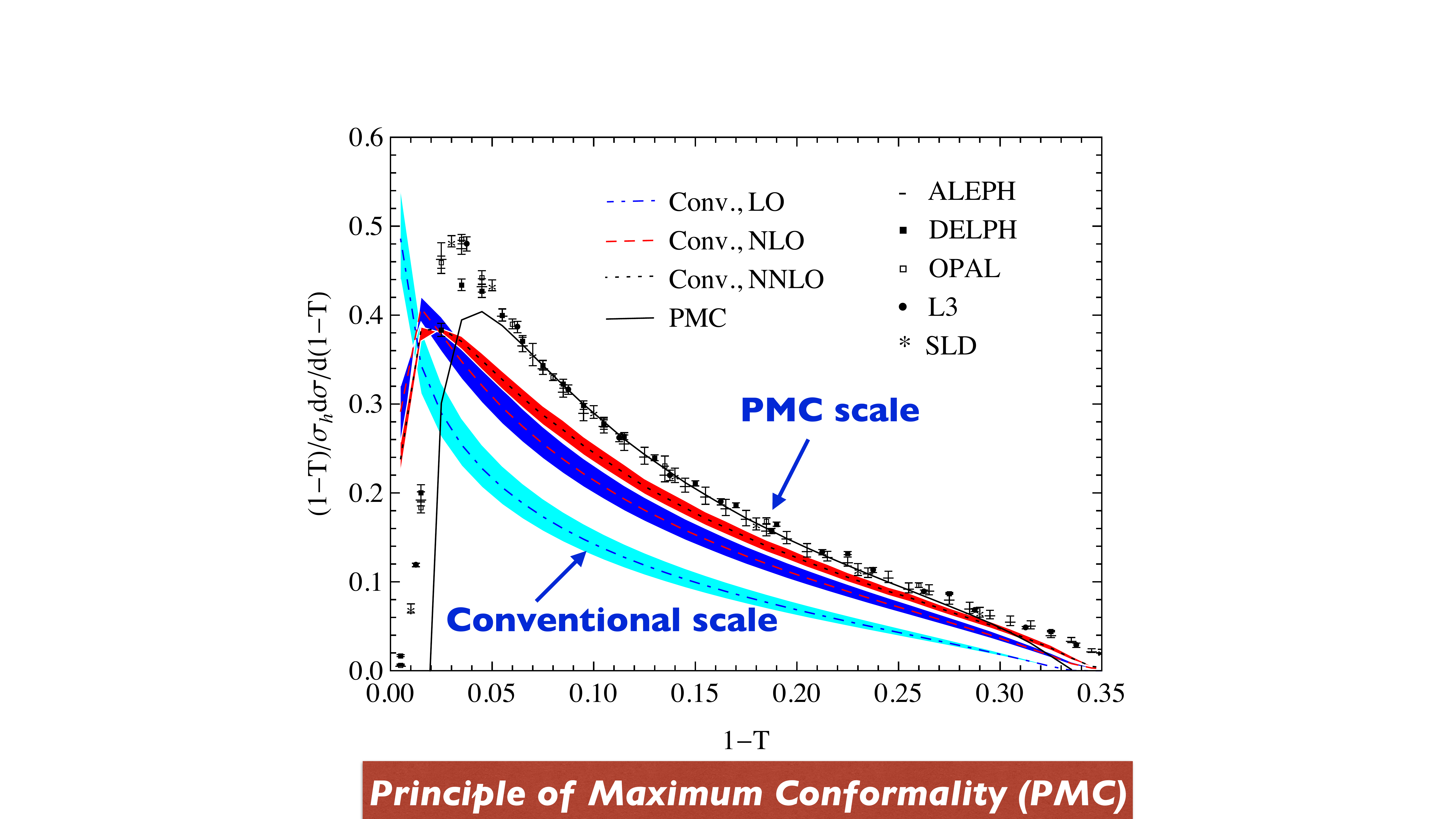}
\end{center}
\caption{Comparison of predictions for the thrust distribution for jet production in $e^+ e^-$ annihilation,  using the PMC to set the pQCD renormalization scale vs. conventional methods.}
\label{Bledslides12.pdf}
\end{figure}

An essential property of renormalizable SU(N)]/U(1) gauge theories, is ``Intrinsic Conformality,"~\cite{DiGiustino:2020fbk}.  It underlies the scale invariance of physical observables and can be used to resolve the conventional renormalization scale ambiguity {\it at every order } in pQCD.  This reflects the underlying conformal properties displayed by pQCD at NNLO, eliminates the scheme dependence of pQCD predictions and is consistent with the general properties of the PMC.   We have also introduced a new method \cite{DiGiustino:2020fbk} to identify the conformal and $\beta$ terms which can be applied either to numerical or to theoretical calculations and in some cases allows infinite resummation of the pQCD series,
The implementation of the PMC$_\infty$ can significantly improve the precision of pQCD predictions; its implementation in multi-loop analysis also simplifies the calculation of higher orders corrections in a general renormalizable gauge theory.
This method has also been used to improve the NLO  pQCD prediction for $t \bar t$ pair production and other processes at the LHC,  where subtle aspects of the renormalization scale of the three-gluon vertex and multi gluon amplitudes, as well as  large radiative corrections to heavy quarks at threshold play a crucial role.  
The large discrepancy of pQCD predictions with  the forward-backward asymmetry measured at the Tevatron is significantly reduced from 3~$\sigma$ to approximately 1~$\sigma.$
The PMC  has also been used to precisely determine the QCD running coupling constant $\alpha_s(Q^2) $  over a wide range of $Q^2$ from event shapes for electron-positron annihilation measured at a single
energy $\sqrt s$ ~\cite{Wang:2019isi}.
The PMC method has also been applied  to a spectrum of LHC processes including Higgs production, jet shape variables, and final states containing a high $p_T$ photon plus heavy quark jets, all of which, sharpen the precision of the Standard Model  predictions.

\section{Is the Momentum Sum Rule Valid for Nuclear Structure Functions? }

Sum rules for deep inelastic lepton-hadron scattering processes are analyzed using the operator product expansion of the forward virtual Compton amplitude, assuming it depends in the limit $Q^2 \to \infty$ on matrix elements of local operators such as the energy-momentum tensor.  The moments of the structure function and other distributions can then be evaluated as overlaps of the target hadron's light-front wave function,  as in the Drell-Yan-West formulae for hadronic form factors~\cite{Brodsky:1980zm,Liuti:2013cna,Mondal:2015uha,Lorce:2011dv}.
The real phase of the resulting DIS amplitude and its OPE matrix elements reflects the real phase of the stable target hadron's wave function.
The ``handbag" approximation to deeply virtual Compton scattering also defines the ``static"  contribution~\cite{Brodsky:2008xe,Brodsky:2009dv} to the measured parton distribution functions (PDF), transverse momentum distributions, etc.  The resulting momentum, spin and other sum rules reflect the properties of the hadron's light-front wave function.
However, final-state interactions which occur {\it after}  the lepton scatters on the quark, can give non-trivial contributions to deep inelastic scattering processes at leading twist and thus survive at high $Q^2$ and high $W^2 = (q+p)^2.$   For example, the pseudo-$T$-odd Sivers effect~\cite{Brodsky:2002cx} is directly sensitive to the rescattering of the struck quark. 
Similarly, diffractive deep inelastic scattering involves the exchange of a gluon after the quark has been struck by the lepton~\cite{Brodsky:2002ue}.  In each case the corresponding DVCS amplitude is not given by the handbag diagram since interactions between the two currents are essential.
These ``lensing" corrections survive when both $W^2$ and $Q^2$ are large since the vector gluon couplings grow with energy.  Part of the phase can be associated with a Wilson line as an augmented LFWF~\cite{Brodsky:2010vs} which do not affect the moments.  

The cross section for deep inelastic lepton-proton scattering $\ell p \to \ell' p' X$
includes a diffractive deep inelastic (DDIS) contribution 
in which the proton remains intact with a large longitudinal momentum fraction $x_F>0.9$
greater than 0.9 and small transverse momentum. The DDIS events, which can be identified with Pomeron exchange in the 
$t$-channel, account for approximately 10\%
of all of the DIS events. 
Diffractive DIS contributes at leading-twist (Bjorken scaling) and is the essential component of the two-step amplitude which causes shadowing and antishadowing of the nuclear PDF ~\cite{Brodsky:2021jmj, Brodsky:2021bkt, Brodsky:2004qa, Schienbein:2007fs}.
It is important to analyze whether the momentum and other sum rules derived from the OPE expansion in terms of local operators remain valid when these dynamical rescattering corrections to the nuclear PDF are included.   The OPE is derived assuming that the LF time separation between the virtual photons in the forward virtual Compton amplitude 
$\gamma^* A \to \gamma^* A$  scales as $1/Q^2$.
However, the propagation  of the vector system $V$ produced by the diffractive DIS interaction on the front face and its inelastic interaction with the nucleons in the nuclear interior $V + N_b \to X$ are characterized by a longer LF time  which scales as $ {1/W^2}$.  Thus the leading-twist multi-nucleon processes that produce shadowing and antishadowing in a nucleus are evidently not present in the $Q^2 \to \infty$ OPE analysis.

Thus, when one measures DIS, one automatically includes the leading-twist Bjorken-scaling DDIS events as a contribution to the DIS cross section, whether or not the final-state proton is explicitly detected.  In such events, the missing momentum fraction 
in the DDIS events could be misidentified with the light-front momentum fraction carried by sea quarks or gluons in the proton's Fock structure. The underlying QCD Pomeron-exchange amplitude which produces the DDIS events thus does not obey the operator product expansion nor satisfy momentum sum rules -- the quark and gluon distributions measured in DIS experiments will be misidentified, unless the measurements explicitly exclude the DDIS events~\cite{Brodsky:2019jla,Brodsky:2021bkt}

The Glauber propagation  of the vector system $V$ produced by the diffractive DIS interaction on the nuclear front face and its subsequent  inelastic interaction with the nucleons in the nuclear interior $V + N_b \to X$ occurs after the lepton interacts with the struck quark.  
Because of the rescattering dynamics, the DDIS amplitude acquires a complex phase from Pomeron and Regge exchange;  thus final-state  rescattering corrections lead to  nontrivial ``dynamical" contributions to the measured PDFs; i.e., they involve the physics aspects of the scattering process itself~\cite{Brodsky:2013oya}.  The $ I = 1$ Reggeon contribution to diffractive DIS on the front-face nucleon leads to flavor-dependent antishadowing~\cite{Brodsky:1989qz,Brodsky:2004qa}.  This could explain why the NuTeV charged current measurement $\mu A \to \nu X$ scattering does not appear to show antishadowing
 in contrast to deep inelastic electron nucleus scattering~\cite{Schienbein:2007fs}.
Again, the corresponding DVCS amplitude is not given by the handbag diagram since interactions between the two currents are essential to explain the physical phenomena.

It should be emphasized  that shadowing in deep inelastic lepton scattering on a nucleus  involves  nucleons at or near the front surface; i.e, the nucleons facing the incoming lepton beam. This  geometrical orientation is not built into the frame-independent nuclear LFWFs used to evaluate the  matrix elements of local currents.  Thus the dynamical phenomena of leading-twist shadowing and antishadowing appear to invalidate the sum rules for nuclear PDFs.  The same complications occur in the leading-twist analysis of deeply virtual Compton scattering $\gamma^* A \to \gamma^* A$ on a nuclear target.

\section{Summary}
Light-Front Hamiltonian theory provides a causal, frame-independent, ghost-free nonperturbative formalism for analyzing gauge theories such as QCD. 
Remarkably, LF theory in 3+1 physical space-time is holographically dual to five-dimensional AdS space, if one identifies the LF radial variable $\zeta$ with the fifth coordinate $z$ of AdS$_5$.  If the metric of the conformal AdS$_5$ theory is modified by  a dilaton of the form $e^{+ \kappa^2 z^2}$, one obtains an analytically-solvable  Lorentz-invariant color-confining LF Schr\"odinger equations for hadron physics.  The parameter $\kappa$ of the dilaton becomes the fundamental mass scale of QCD, underlying the color-confining potential of the LF Hamiltonian and the running coupling $\alpha_s(Q^2)$ in the nonperturbative domain.  When one introduces super-conformal algebra, the result is ``Holographic LF QCD" which not only predicts a unified Regge-spectroscopy of mesons, baryons, and tetraquarks, arranged as supersymmetric 4-plets, but also the hadronic LF wavefunctions which underly form factors, structure functions,  and  other dynamical phenomena.  In each case, the quarks and antiquarks cluster in hadrons as $3_C$ diquarks, so that mesons, baryons and tetraquarks all obey a two-body $3_C - \bar 3_C$ LF bound-state equation.  Thus tetraquarks are compact hadrons, as fundamental as mesons and baryons.
``Holographic LF QCD" also leads to novel phenomena such as the color transparency of hadrons produced in 
hard-exclusive reactions traversing a nuclear medium and asymmetric intrinsic heavy-quark distributions $Q(x) \ne \bar Q(x)$, appearing at high $x$ in the non-valence higher Fock states of hadrons.

\section*{Acknowledgements}

Contribution to 
the Proceedings of  the 24th Workshop, ``What Comes Beyond
the Standard Models", Bled, July 3. -- 11., 2021.
I am grateful to all of my collaborators, especially Guy  de T\'eramond, Hans Guenter Dosch, Cedric Lorc\'e, Maria Nielsen,  Tianbo Liu, Sabbir Sufian,  and Alexandre Deur, for their collaboration on light-front holography and its implications.
This work is supported by the Department of Energy, Contract DE--AC02--76SF00515. SLAC-PUB-17634

%SLAC-PUB-XXXXX.

%% The bibliography section

\end{document}